\documentclass[acmtog]{acmart}
\usepackage{booktabs} % For formal tables
\usepackage{caption}
\usepackage{subcaption}
\usepackage{xfrac}

\usepackage[ruled]{algorithm2e} % For algorithms
\usepackage{xcolor}
\usepackage{svg}

\SetAlFnt{\small}
\SetAlCapFnt{\small}
\SetAlCapNameFnt{\small}
\SetAlCapHSkip{0pt}

\newcommand{\mb}{\boldsymbol}

\newcommand{\bb}{\mb{b}}

\newcommand{\ee}{\mb{e}}
\newcommand{\FF}{\mb{F}}
\newcommand{\ff}{\mb{f}}
\renewcommand{\gg}{\mb{g}}
\newcommand{\ii}{\mb{i}}
\newcommand{\nn}{\mb{n}}
\newcommand{\uu}{\mb{u}}
\newcommand{\VV}{\mb{V}}
\renewcommand{\vv}{\mb{v}}
\newcommand{\XX}{\mb{X}}
\newcommand{\xx}{\mb{x}}
\renewcommand{\ss}{\mb{s}}
\newcommand{\zz}{\mb{z}}

\newcommand{\dx}{\Delta x}
\newcommand{\dt}{\Delta t}

\acmJournal{TOG}
\begin{document}
% Title portion
\title{A Thermomechanical Hybrid Incompressible Material Point Method}
\begin{teaserfigure}
    \centering
    \includegraphics[width=\textwidth]{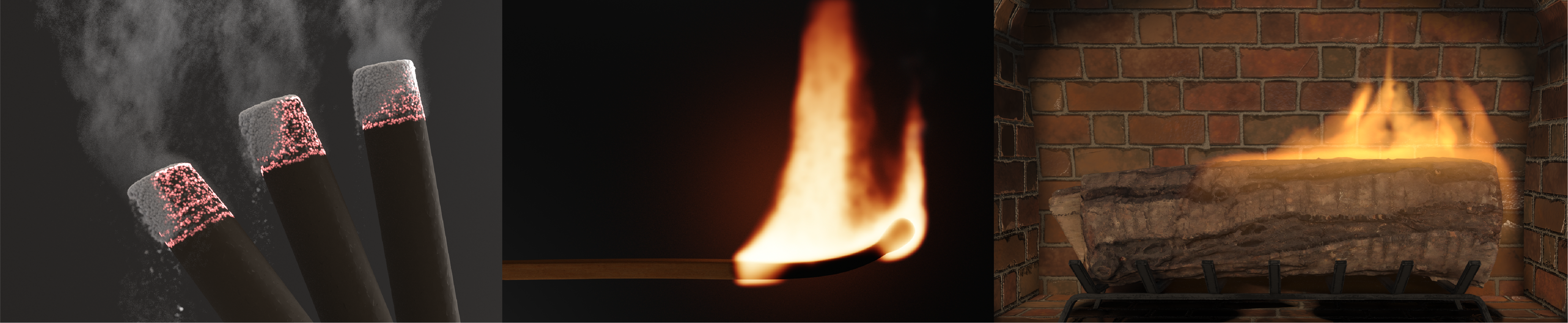}
    \caption{Our method is capable of simulating the combustion of solid objects and the resulting fire, smoke, and ash.}
\end{teaserfigure}

% DO NOT ENTER AUTHOR INFORMATION FOR ANONYMOUS TECHNICAL PAPER SUBMISSIONS TO SIGGRAPH 2019!
\author{Victoria Kala}
%\orcid{1234-5678-9012-3456}
\affiliation{%
  \institution{University of Utah}
  \streetaddress{155 S 1400 E, RM 233}
  \city{Salt Lake City}
  \state{UT}
  \postcode{84112}
  \country{USA}
}
\email{victoria.kala@utah.edu}

\author{Jingyu Chen}
\affiliation{%
  \institution{University of California, Los Angeles}
  \city{Los Angeles}
  \state{CA}
  \postcode{90095}
  \country{USA}
}
\email{chenjy@g.ucla.edu}

\author{David Hyde}
\affiliation{%
 \institution{Vanderbilt University}
  \streetaddress{1400 18th Ave S}
  \city{Nashville}
  \state{TN}
  \postcode{37212}
  \country{USA}
}
\email{david.hyde.1@vanderbilt.edu}

\author{Alexey Stomakhin}
\affiliation{%
  \institution{Wētā FX}
  \streetaddress{PO Box 15186 }
  \city{Miramar}
  \state{Wellington}
  \postcode{6243}
  \country{New Zealand}
}
\email{st.alexey@gmail.com}

\author{Joseph Teran}
\affiliation{%
  \institution{University of California, Davis}
  \streetaddress{1 Shields Ave}
  \city{Davis}
  \state{CA}
  \postcode{95616}
  \country{USA}
}
\email{jteran@math.ucdavis.edu}

\begin{abstract}
We present a novel hybrid incompressible flow/material point method solver for simulating the combustion of flammable solids.
Our approach utilizes a sparse grid representation of solid materials in the material point method portion of the solver and a hybrid Eulerian/FLIP solver for the incompressible portion.
We utilize these components to simulate the motion of heated air and particulate matter as they interact with flammable solids, causing combustion-related damage.
We include a novel particle sampling strategy to increase Eulerian flow accuracy near regions of high temperature.
We also support control of the flame front propagation speed and the rate of solid combustion in an artistically directable manner.
Solid combustion is modeled with temperature-dependent elastoplastic constitutive modeling.
We demonstrate the efficacy of our method on various real-world three-dimensional problems, including a burning match, incense sticks, and a wood log in a fireplace.
\end{abstract}





\maketitle

\section{Introduction}
\label{sec:intro}

From films such as \textit{Elemental} (2023) to scenes from \textit{The Dark Knight} (2008), physical phenomena such as fire and burning are essential in the modern palette of computer-generated special effects.
However, burning, combustion, and related effects are highly complex, and accurately simulating such phenomena---particularly when artistic direction is desired---remains challenging, so much so that recent films such as \textit{Star Wars: The Last Jedi} (2017) relied on practical effects for its pivotal tree burning scene.

When an object undergoes combustion, several areas of chemistry, physics, and mechanics suddenly interact.
Chemical reactions dictate how a fuel source (including any combustible material) reacts to become fire or soot.
Systems of partial differential equations describe the motion of flames, smoke, and ash through the surrounding air.
Material parameters of the object being burned dictate behavior such as drying, compression, and fracture.
All these phenomena need to be modeled or approximated when building a simulation system for these types of thermomechanical problems.

We propose an approach based on the material point method (MPM) \citep{sulsky:1994:history-materials,jiang:2016:course} to attack these simulation challenges.
MPM is a hybrid particle-grid method, and we design our method to leverage the advantages of both particle-based discretizations, e.g., particles providing additional spatial adaptivity in regions where the computational background grid is under-resolved, and grid-based discretizations, e.g., being able to use a standard incompressible fluid solver to evolve flames and smoke under the incompressible Euler equations.
Moreover, from a mathematical perspective, MPM naturally supports dynamic topology changes (see, e.g., \citet{chen:2021:momentum}), and from a computational perspective, optimized implementations of MPM have been demonstrated to scale remarkably efficiently on modern hardware \citep{wang:2020:massively}.

The contributions of our work may be summarized as follows:
\begin{itemize}
    \item A sparse grid MPM implementation for burning solids, coupled with a hybrid Eulerian / FLIP incompressible fluid solver for gases.
    \item Temperature-dependent elastoplastic constitutive modeling that simulates visually plausible solid deformation due to combustion.
    \item A novel smoke particle-sampling strategy to increase Eulerian flow accuracy near regions of high temperature.
    \item A novel ignition method that supports control of the flame front propagation speed and rate of solid combustion.
\end{itemize}

After discussing related work in Section \ref{sec:related-work}, we present our method in Sections \ref{sec:governing-equations} and  \ref{sec:sim-framework}.
We showcase several practical, three-dimensional examples of our approach in Section \ref{sec:examples}, and discuss limitations and future work in Section \ref{sec:conclusions}.

\section{Related Work}
\label{sec:related-work}

Significant progress has been achieved in simulating fluid mechanics, fire dynamics, and combustion processes.
We highlight the algorithms described in \citet{bridson2015fluid} and \citet{guendelman2005coupling}, which are the basis of our incompressible solver in the present work.
Our solver encapsulates the intricacies of various phenomena, including smoke, which can be modeled with incompressible flows (see, e.g., \citet{fedkiw:2001:visual}). 
Beyond smoke, \citet{nguyen2002physically} presented a physically based model for simulating fire.
A method for animating suspended particle explosions, focusing on simulating the dynamics of particulate matter within explosive events, was proposed by \citet{feldman2003animating}.
\citet{hong2007wrinkled} investigated the formation of wrinkled flames.
\citet{horvath2009directable} developed a directable, high-resolution simulation of fire on the GPU, enabling realistic and interactive fire animations for various applications.
``Loki'' was developed as a unified multiphysics simulation framework for production, offering comprehensive capabilities for simulating fire and combustion alongside other physical phenomena \citep{lesser2022loki}.
\citet{nielsen2022physics} further advanced physics-based gaseous combustion simulation with an approach that includes modeling of real-world fuels, soot formation and oxidation, radiative heating, and other thermodynamic effects.
Fire and combustion algorithms similar to those in \citet{nielsen2022physics} were used for the elemental phenomena in \textit{Avatar: The Way of Water} (2022) \citep{edholm:2023:fire}.

While there has been extensive research regarding simulation for fire and combustion, there remains room for further exploration into visually plausible combustion of solids, a gap we aim to address. 
Level sets are often used to simulate the decomposition of burning objects, as seen in \citet{melek:2003:interactive,melek:2004:modeling,losasso:2006:melting,riensche:2009:modeling}.
Combustion for thin objects such as paper and cloth have also been studied in \citet{moidu:2004:animating,melek:2007:driving,celly:2015:smoldering,larboulette:2013:burning}. 
There has also been interest in simulating combustion of trees and spread of wildfires, as seen in \citet{pirk2017interactive,hadrich2021fire}.
\citet{jo:2019:lattice} proposed a hybrid method for burning solid interactions that can perform smoldering; however, it ignored other solid deformation, such as the burning solid fracturing or transitioning to ash.

As addressed in Section \ref{sec:intro}, MPM is a compelling choice for modeling solid combustion due to its ability to handle diverse materials and topological changes. 
Solid combustion often entails phenomena such as shrinking, fracturing, or transitioning from a solid to granular material such as ash, and recent contributions in the realm of MPM within computer graphics have addressed applications to these properties.
For instance, MPM was used to model granular materials such as snow and sand \citep{daviet:2016:smp,klar:2016:des,yue:2018:hmpm,stomakhin:2013:snow}.
\citet{wang:2019:fracture} and \citet{fan2022simulating} developed methods for simulating fracture using material points. Furthermore,
heat transfer and phase change have been modeled with MPM in  \citet{stomakhin:2014:augmented-mpm,gao:2017:ampm,chen:2021:momentum}. 

Solid simulated in the present work may undergo phase transition from solid to a granular state, such as ash.
Additionally, combustion will be largely dictated by the surrounding fluids such as fire, smoke, and air.
Hence it is important our method be capable of modeling multispecies solids as well as coupling with incompressible fluids.
MPM has been used to couple multiple materials and simulate multispecies mixtures, as seen in \citet{abe:2014:mpm,bandara:2015:csd,bandara:2016:mpm,tampubolon:2017:wetsand}.

\section{Governing Equations}
\label{sec:governing-equations}

\subsection{Continuous Kinematics}

As in works such as \citet{gonzalez:2008:continuum,hyde:2020:tension}, we make the assumption that objects can be represented as a continuum with particles throughout the material.
We let $\Omega^t \subset \mathbb{R}^d$ denote the domain of the material at time $t$, $0 \leq t \leq T$, in $d$-dimensional space ($d = 2, 3$).
Within this domain, continuum particles $\XX \in \Omega^t$ have trajectories defined by a flow map $\boldsymbol{\phi}: \Omega^0 \times [0,T] \to \mathbb{R}^d$.
We let $\xx = \boldsymbol{\phi}(\XX, t)$ denote particle positions at time $t$.

Within this notation, the velocity of the material can be obtained as the temporal partial derivative of the flow map, $\VV = \dfrac{ \partial \boldsymbol{\phi} }{\partial t}$.
The deformation gradient of the material is given by the corresponding spatial derivative, $\FF = \dfrac{ \partial \boldsymbol{\phi} }{\partial \XX}$.
We let $J = \text{det}(\FF)$ be the determinant of the deformation gradient.
$J$ is useful for tracking the amount of volumetric dilation at a given material point.
%We refer to \citet{gonzalez:2008:continuum,hyde:2020:tension} for further details of 

It is often more convenient to write quantities using their time $t$ configuration in $\Omega^t$ rather than in $\Omega^0$.
We can take Lagrangian variables such as $\FF$, $J$, and $\boldsymbol{\phi}$ and perform a change of variables to produce their Eulerian representations based on $\Omega^t$.
For instance, we can write the Eulerian velocity of the material as $\vv\left(\xx, t \right) = \VV \left( \boldsymbol{\phi}^{-1} (\xx, t), t \right)$, where $\boldsymbol{\phi}^{-1}$ is the inverse of the flow map (see \citet{gonzalez:2008:continuum}).

\subsection{Conservation of Mass and Momentum}

Solid materials simulated in this paper are governed by conservation of mass and momentum.
In Eulerian form, these two constraints are written as
\begin{align}\label{eq:balance}
\frac{D\rho}{Dt}=-\rho\nabla\cdot\vv, \rho\frac{D\vv}{Dt}&=\nabla \cdot \boldsymbol{\sigma} + \rho \gg, \ \xx\in\Omega^t.
\end{align}
Here, $\rho$ represents material density, $\vv$ is velocity, $\boldsymbol{\sigma}$ is the Cauchy stress, and $\gg$ is gravitational acceleration.
$D/Dt$ denotes the material derivative.
Solids act as boundary conditions for the fluid solver.

\subsection{Incompressible Flow}

We model fluids such as smoke via the inviscid, incompressible Euler equations, which are given by 
\begin{align}
    \dfrac{\partial \uu}{\partial t} + (\uu \cdot \nabla) \uu + \dfrac{1}{\rho} \nabla p = \ff, 
    \nabla \cdot \uu = 0 ,
\end{align}
where $\uu$ is the fluid velocity and $\ff$ are body forces.
We solve these equations via the standard Chorin projection scheme \citep{chorin:1967:numerical}, which splits the equations into an advection step and application of body forces, followed by a pressure projection to enforce incompressibility:
\begin{align}
    \dfrac{D \uu}{Dt} = 0 & \quad \text{(advection)} \label{eq:advection} \\
    \dfrac{\partial \uu}{\partial t} = \ff & \quad \text{(body forces)} \label{eq:body_forces} \\
    \dfrac{\partial \uu}{\partial t} + \dfrac{1}{\rho} \nabla p = 0 ~ \text{such that} ~ \nabla \cdot \uu = 0 & \quad  \text{\small{(pressure/incompressibility)}} \label{eq:pressure}
\end{align}

\subsection{Thermal Diffusion}

To capture thermodynamic effects in a solid or fluid, thermal diffusion can be modeled by
\begin{equation}
    \rho c_p \dfrac{DT}{Dt} = K \nabla^2 T + H . \label{diffusion_solve}
\end{equation}
Here, $c_p$ is the specific heat capacity, $T$ is the temperature, $K$ is thermal conductivity, and $H$ is a source function (set to $0$ in this paper, as we alternatively increase temperature in an ignition step as described in Section \ref{section:update_temperature}).
Since the heat capacity and thermal conductivity will have different values for a burning solid and surrounding air, we model $K$ and $c_p$ with Heaviside functions based on a level set $\phi(\xx)$ generated for the material on each time step.
For example, we can model $K$ as
\begin{equation}
K(\xx) = \begin{cases} K_{\text{air}} & \phi(\xx) > 0 \\ K_{\text{solid}} & \phi(\xx) < 0 \end{cases} ,
\end{equation}
where $\phi(\xx) > 0$ is outside the solid and $\phi(\xx) < 0$ is inside the solid.
The specific heat capacity $c_p$ is modeled analogously.

\subsection{Solid Constitutive Models}
\label{sec:cons-models}

For most solids simulated in this paper, we use the so-called fixed-corotated constitutive model \citep{jiang:2016:course} from \citet{stomakhin:2012:invertible},
\begin{equation}
    \hat{\boldsymbol{\psi}} = \mu \sum_i \left( \sigma_i - 1 \right)^2 + \frac{\lambda}{2} \left( J - 1 \right)^2 ,
\end{equation}
where $\hat{\boldsymbol{\psi}}$ is energy density, $\mu$ and $\lambda$ are the hyperelastic Lam\'e coefficients, and the $\sigma_i$ are the singular values of $\FF$.

Furthermore, for burnt particles in the incense example (Section \ref{sec:incense}), we use the St.\ Venant-Kirchhoff hyperelastic model with Hencky strains,
\begin{equation}
    \hat{\boldsymbol{\psi}} = \mu \text{tr}\left( \boldsymbol{\eta}^2 \right) + \frac{\lambda}{2} \text{tr}(\boldsymbol{\eta})^2 ,
\end{equation}
where $\eta$ is the Hencky strain tensor $\eta = \frac{1}{2} \text{ln}\left(\FF \FF^T \right)$ \citep{xiao:2005:hencky}.
Plasticity is also modeled for burnt particles in that example, using the Drucker-Prager model \citep{drucker:1952:elastoplast,klar:2016:des}.We selected these constitutive models based on their proven efficacy in accurately representing the mechanical behavior of both elastic solids and elastoplastic granular materials, see, e.g., \citet{klar:2016:des}.

\section{Simulation Loop and Discretization} \label{sec:sim-framework}

%\begin{figure}
%    \centering
%    \includegraphics[scale=0.7]{figs/overview.png}
%    \caption{An overview of our algorithm.}
%    \label{fig:overview}
%\end{figure}

Our algorithm begins by performing an initialization step (Section \ref{sec:initialization}).
After initialization, the simulation loop of our method consists of four main steps for each time step:

%Here we describe the discretization details of algorithm. We outline each step required to advance one time step in the simulation other than the initialization step. We can think of this process as updating the state from time $t^n$ to time $t^{n+1}$. We discuss the specific details of the discretization each in step in the algorithm in the following subsections, which can be summarized as:

\begin{enumerate}
    %\item Initialization (performed once, Section \ref{sec:initialization})
    \item \textbf{MPM Step} (Section \ref{sec:MPM_step}). We transfer particle information to a background Eulerian grid and update the forces on the grid. In our force update, we can update the constitutive model and apply isotropic and anisotropic shrinking. We also sample smoke particles. Finally, we transfer updated quantities from the grid back to particles. % from particles back to grid.
    %\begin{enumerate}
    %    \item Transfer from particles to grid (Section \ref{sec:P2G})
    %    \item Calculate forces on grid (Section \ref{sec:grid_update})
    %    \item Update Phases (Section \ref{sec:update_phases})
    %    \item Isotropic Shrinking (Section \ref{section:isotropic_shrinking})
    %    \item Anisotropic Shrinking (Section \ref{section:anisotropic_shrinking})
    %    \item Resample Smoke (Section \ref{section:resample_smoke})
    %    \item Transfer from particles to grid (Section \ref{sec:G2P})
    %\end{enumerate}
    \item \textbf{Incompressible Step} (Section \ref{section:incompressible_step}). We identify boundary conditions from the solid and apply them to the fluid. We solve on the incompressible grid using a standard Chorin splitting approach. We then apply particle advection using our sampled smoke particles.
    %\begin{enumerate}
    %    \item Apply Fluid Boundary Conditions (Section \ref{section:apply_fluid_bcs})
    %    \item Advection (Section \ref{section:incompressible_advection})
    %    \item Nodes to Face Velocities (Section \ref{section:nodes_to_face_velocities})
    %    \item Apply Forces (Section \ref{section:apply_forces})
    %    \item Pressure Solve (Section \ref{section:pressure_solve})
    %    \item Apply Pressure Gradient (Section \ref{section:apply_pressure_gradient})
    %    \item Extrapolation (Section \ref{section:extrapolation})
    %    \item Face to Node Velocities (Section \ref{section:face_to_node_velocities})
    %    \item Particle Advection (Section \ref{section:particle_advection})
    %\end{enumerate}
    \item \textbf{Temperature Step} (Section \ref{section:temperature_step}). We advect temperatures using the fluid velocity from the incompressible grid. We transfer temperatures from the MPM object to the associated background MPM grid. We merge the MPM and fluid temperatures to a single temperature grid, then solve the diffusion equation on this grid. We extrapolate the fluid temperatures where the solid was so we can have reliable advection for temperature in the next step. We transfer the updated temperature from the grid back to the particles.
    %\begin{enumerate}
    %    \item Advection %(Section \ref{section:temperature_advection})
    %    \item Splat Temperature to Grid (P2G) %(Section \ref{section:temperature_p2g})
    %    \item Merge MPM and Fluid Temperatures %(Section \ref{section:merge_temperatures})
    %    \item Temperature Solve %(Section \ref{section:temperature_solve})
    %    \item Temperature Extrapolation %(Section \ref{section:temperature_extrapolation})
    %    \item Splat Grid Temperature to Particles (G2P) %(Section \ref{section:temperature_g2p})
    %\end{enumerate}
    \item \textbf{Ignition Step} (Section \ref{section:ignition_step}). We provide a novel method for simulating the burning flame front on the surface of an object. We introduce a fuel level and temperature update.
    Temperature and fuel level determine a particle's burning state and lead to igniting neighboring particles. 
    %\begin{enumerate}
    %    \item Update fuel level %(Section \ref{section:update_fuel_level})
    %    \item Update temperature %(Section \ref{section:update_temperature})
    %    \item Determine particle burning state %(Section \ref{section:determine_burning_state})
     %   \item Ignite neighboring particles %(Section \ref{section:ignite_neighboring_particles})
    %\end{enumerate}
\end{enumerate}

We note that our implementation uses several grid objects: an MPM grid, an incompressible velocity grid, a pressure grid, a temperature grid, and a level set grid.
The MPM, pressure, temperature, and level set grid nodes are located at the cell centers of the incompressible grid.
The MPM grid uses quadratic interpolation, while the others use linear interpolation.
These grid objects can largely be viewed as different arrays on the same grid.
%An illustration of the grids is shown in figure \ref{fig:incompressible_pressure_grids}.

\subsection{Initialization} \label{sec:initialization}

%\begin{figure}
%    \centering
%    \includegraphics[scale=0.5]%{figs/incompressible_pressure_grids.png}
%    \caption{The incompressible (blue) and cell center (orange) grids for $N = 3$.}
%    \label{fig:incompressible_pressure_grids}
%\end{figure}

We keep track of several properties of each MPM particle, such as mass, velocity, temperature, temperature gradient, phase, fuel level, burning status, and time to burn.
For the smoke particles (Section \ref{sec:MPM_step}), we track mass, fuel level, temperature, temperature gradient, and velocity.
We also store temperature and density for the surrounding incompressible fluid.
All of these quantities are initialized according to the prescribed initial conditions for the given problem.

For simulations that require a mesh for post-process mesh fracturing (Section \ref{sec:fracture}), we generate a tetrahedral mesh of the object, using tools such as Tetwild \citep{hu:2018:tetwild} or SideFX's Houdini Remesh.
We generate MPM particles at the vertices in the tetrahedral mesh.

Additionally, we can specify a point or region where ignition initially occurs.
This can be chosen, for instance, by identifying the closest particle on the boundary of a solid to a desired point, or by manually selecting a set of particles to initially ignite.

\subsection{MPM Step}
\label{sec:MPM_step}

%We use a sparse MPM grid. We use can use either implicit method (incense and log example) or quasistatic (match example). TOOD: More about sparse MPM? Jingyu?
Our method supports different time integration strategies, including explicit MPM, implicit MPM, and quasi-static MPM.
We use implicit MPM for the examples in Sections \ref{sec:log} and \ref{sec:incense}, and quasi-static MPM for the example in Section \ref{sec:match}.
Particle quantities are transferred to the background Eulerian grid using APIC \citep{jiang:2015:apic}.
Most quantities are transferred using quadratic APIC, while linear APIC is used for temperature to improve sharpness.

As solids burn and combust, they undergo deformation.
Sections \ref{section:isotropic_shrinking}--\ref{sec:update_phases} provide three methods that can be used to simulate this deformation during the MPM stress update.

 %\subsubsection{Transfer from particles to grid} \label{sec:P2G}

% We transfer mass and momentum to the grid in the standard APIC (TODO: equations necessary?):
 %\begin{align}
 %    m_{\ii} & = \sum_p w_{\ii p} m_p \\
 %    m_{\ii} \vv_{\ii} & = \sum_p w_{\ii p} m_p (\vv_p + \BB_p (\DD_p)^{-1}(\xx_{\ii} - \xx_p))
 %\end{align}

 %\subsubsection{Calculate forces on grid.} \label{sec:grid_update}

% We  use implicit method or quasisatic method to update the forces on the grid. We update in the usual way for elastic or elastoplastic materials. No new contribution.

\subsubsection{Isotropic Shrinking}
\label{section:isotropic_shrinking}

We can model shrinking of burning objects by scaling the rest state.
We do this by multiplying $\FF_p$, the elastic deformation gradient for particle $p$, by $1 + c \Delta t$, where $c$ is a coefficient based on the temperature:
\begin{equation}
c = \dfrac{c_\text{shrink} - 1}{T_\text{max} - T_\text{evap}} \left(T_p - T_\text{evap}\right) + 1 . \label{shrink_scale}
\end{equation}
Here, $c_\text{shrink}$ is a shrinking scale constant greater than but close to 1 (e.g., $1 + \varepsilon$ for small $\varepsilon$), $T_\text{max}$ is the maximum temperature of the MPM object, $T_\text{evap}$ is the evaporation temperature, and $T_p$ is the temperature of particle $p$.
If $c_\text{shrink}$ is chosen exactly equal to 1, then no shrinking will occur.
Isotropic shrinking is demonstrated in the match example in Section \ref{sec:match}.

\subsubsection{Anisotropic Shrinking} \label{section:anisotropic_shrinking}
Alternatively, we can shrink an object about its central axis.
For instance, wood is an anisotropic material and shrinks in the radial direction and longitudinal directions.

Consider the cylinder $y^2 + z^2 = r^2$.
Using the change of coordinates $x = x, y = r \sin{\theta}, z = r\cos{\theta}$, where $\tan{\theta} = \dfrac{y}{z}$, we can write a new orthonormal basis:
\begin{align}
    \ee_x = (1, 0, 0)^T,\quad 
    \ee_\theta = \left( 0, \dfrac{z}{r}, -\dfrac{y}{r} \right)^T,\quad 
    \ee_r = \left(0, \dfrac{y}{r}, \dfrac{z}{r} \right)^T .
\end{align}
We can write the deformation gradient $\FF$ in terms of this new basis:
\begin{equation}
\FF = \ee_x \otimes \uu_1 + \ee_\theta \otimes \uu_2 + \ee_r \otimes \uu_3 ,
\end{equation}
where $\otimes$ denotes the outer product. Expanding the outer products yields
\begin{equation} 
    \FF = \left( \begin{matrix}
    u_{11} & u_{12} & u_{13} \\
    (u_{21}z + u_{31}y)/r & (u_{22}z + u_{32}y)/r & (u_{23}z + u_{33}y)/r \\
    (u_{31}z -u_{21}y)/r & (u_{32}z -u_{22}y)/r & (u_{33}z -u_{23}y)/r 
    \end{matrix} \right) .
\end{equation}
We can solve for $u_{ij}$ in terms of $F_{ij}, r, y, z$:
\begin{align}
    %u_{11} & = F_{11} \\
    %u_{12} & = F_{12} \\
    %u_{13} & = F_{13} \\
    %u_{21} & = (F_{21}z - F_{31}y)/r \\
    %u_{22} & = (F_{22}z - F_{32}y)/r \\
    %u_{23} & = (F_{23}z - F_{33}y)/r \\
    %u_{31} & = (F_{21}y + F_{31}z)/r \\
    %u_{32} & = (F_{22}y + F_{32}z)/r \\
    %u_{33} & = (F_{23}y + F_{33}z)/r
    u_{1j} = F_{1j}, \quad
    u_{2j} = (F_{2j}z - F_{3j}y)/r, \quad
    u_{3j} = (F_{2j}y + F_{3j}z)/r .
\end{align}
We can then shrink a cylindrical object longitudinally or radially by scaling $u_{11}$ or  $u_{22}$, respectively, by a constant of the form $c_{\text{shrink}} = 1 + \varepsilon$ for small $\varepsilon$. This method was used in our log example in Section \ref{sec:log}.

\subsubsection{Update Constitutive Models} 
\label{sec:update_phases}

We can update the constitutive model of a particle depending on its burning state, similar to the phase change approach used in \citet{chen:2021:momentum}.
In our incense example in Section \ref{sec:incense}, the constitutive model of burnt particles are changed to be St.\ Venant-Kirchhoff with Hencky strains, and a Drucker-Prager model is applied for elasticity (see Section \ref{sec:cons-models}).

\subsubsection{Smoke Particle Sampling} \label{section:resample_smoke}

In order to achieve greater visual realism, we sample additional particles, representing smoke, on the boundary of a solid as it burns.
We construct a particle level set $\phi$ for all combustible MPM particles and use the boundary particles on the zero isocontour of the level set as the location for sampling these smoke particles (see \cite{chen:2021:momentum} for details of computing boundary particles).
A particle radius of $\frac{\sqrt{d}}{2}\dx$ with $d = 2, 3$ is used for building the level set, where $\Delta x$ is the width of a (uniform) grid cell in the background Eulerian grid.
For each burning MPM particle $\xx_p^n$, we find its closest boundary particle $\bb^n$ and sample $N^s$ particles.
The new smoke particles are sampled at $\ss_r^n = \bb^n + c\dx$, where $c$ is a random value in $[-0.5, 0.5]$ and $0 \leq r \leq N^s-1$.
The smoke particles are given constant mass and zero initial velocity.
Each smoke particle's initial temperature is set to a constant ignition threshold $T_{\text{ignition}}$ to mimic the effect that smoke particles should only be emitted when a particle is burning (and hence is at least at temperature $T_{\text{ignition}}$).

% We construct a level set $\phi$ for our surface using TODO. We build smoke resampling positions using TODO. At each of the smoke resampling positions, we can set the number of smoke particles we want to sample at that location. We initially set the velocity to 0. 

%\subsubsection{Transfer from grid to particles} \label{sec:G2P}

%We transfer from grid to particles using APIC (TODO: equations necessary?).

\subsection{Incompressible Step} \label{section:incompressible_step}

%\begin{figure}
%    \centering
%    \includegraphics[scale=0.85]{figs/MAC_grid_notation.png}
%    \caption{Standard MAC grid notation}
%    \label{fig:MAC_grid_notation}
%\end{figure}

We leverage a standard MAC grid discretization for storing fluid quantities \citep{harlow:1965:mac}.
However, we use a nodal based fluid solver as in \citet{guendelman2005coupling}, where the velocities are stored at the grid nodes of the incompressible grid, i.e., at $\uu_{i\pm1/2,j\pm1/2,k\pm1/2}$ in three spatial dimensions.
We apply solid Neumann boundary conditions from the MPM object to the fluid as described in Section \ref{section:apply_fluid_bcs}.
As popularized by \citet{chorin:1967:numerical}, the incompressible step involves an advection step (Section \ref{section:incompressible_advection}), application of external/body forces (Section \ref{section:apply_forces}), and a projection to enforce incompressibility.
The projection step uses a standard conjugate gradients solver.

\subsubsection{Fluid Boundary Conditions} \label{section:apply_fluid_bcs}

\begin{figure}
    \centering
    \resizebox{0.45\textwidth}{!}{\includesvg{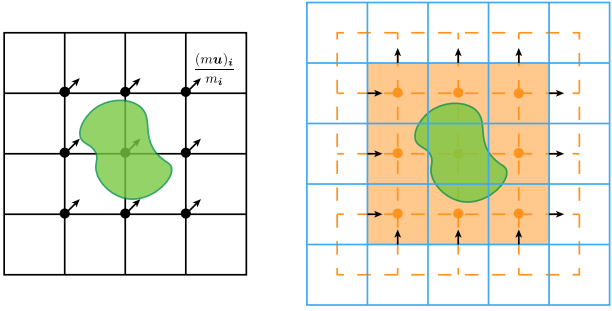}}
    \caption{After transferring mass and momentum from the MPM object (green) to the MPM grid, we identify grid nodes on the MPM grid (black) that have positive mass. We mark solid pressure cells at pressure grid nodes (orange) that correspond to the MPM grid nodes with mass, and use a Neumann velocity boundary condition for the incompressible grid (blue) by dividing the grid momentum by grid mass at those grid nodes. The components of this velocity are then set as the face velocities of the solid pressure cell.}
    \label{fig:apply_fluid_bcs}
\end{figure}

For the domain boundary of the pressure grid (located at the cell centers of the fluid velocity grid), we can either apply Dirichlet or Neumann boundary conditions.
A Dirichlet boundary condition means the pressure $p$ on the boundary is known (usually 0).
A Neumann boundary condition means $\nabla p \cdot \nn$ is known, where $\nn$ is the normal on the boundary.
This is equivalent to knowing the fluid velocity on the face with normal $\nn$.
By default, we set the bottom of the domain to be Neumann with \textbf{0} fluid velocity (i.e., no fluid can leave the bottom of the domain) and the remaining faces of the domain to be Dirichlet with 0 pressure.

Additionally, we can prescribe Neumann boundary conditions based on the MPM solid.
After transferring the MPM mass to the MPM grid, we identify grid nodes with nonzero mass. Grid nodes with nonzero mass are marked as solid pressure cells and given a solid velocity by dividing the grid momentum by the grid mass at that node, i.e. $\dfrac{(m \uu)_{\ii}}{m_{\ii}}$.
The components of this velocity are then set as the face velocities of the solid pressure cell.
These Neumann face velocities are then treated as solid boundary conditions for the fluid.
See Figure \ref{fig:apply_fluid_bcs}.

\subsubsection{Advection} \label{section:incompressible_advection}

%\begin{figure}
%    \centering
%    \includegraphics[scale=0.6]{figs/advection.png}
%    \caption{We interpolate the surrounding node values (blue) of $\xx_{\text{upwind}}$ in our Semi-Lagrangian advection scheme.}
%    \label{fig:advection}
%\end{figure}

We use a particle-assisted advection scheme similar to the narrow-band FLIP method \citep{ferstl:2016:nbflip}.
At each step, we first transfer the momentum of the smoke particles to the background grid in FLIP style using the linear interpolation function $N_{\ii}(\xx_p)$ between particle $\xx_p$ and grid node $\xx_{\ii}$:
\begin{equation}
    m_{\ii} \uu^{\text{FLIP, }n}_{\ii} = \sum_p N_{\ii}(\xx_p^n) m_p \vv_p^n.
\end{equation}
We note that $\uu_i$ is a fluid velocity while $\vv_p^n$ is a particle velocity.
For grid node $\xx_{\ii}$, we update its velocity using 
\begin{equation}
    \uu^A_{\ii} = 
    \begin{cases}
        \uu^{\text{FLIP, }n}_{\ii}, & \text{for } m_{\ii} \neq 0, \\
        \text{semi-Lagrangian}(\xx_{\ii}, \uu_{\ii}^n, \dt), & \text{otherwise}.
    \end{cases}
\end{equation}
We use a standard semi-Lagrangian advection scheme as in \citet{bridson2015fluid} to advect the fluid grid velocities. 
For every grid node $\xx_{\ii}$, we calculate the upwind position $\xx^{\text{upwind}}_{\ii}$ using a third-order Runge-Kutta method.
If the upwind value is not in the fluid grid or $\xx_{\ii}$ is a domain boundary node, we set the advected velocity at grid node $\ii$ directly from the fluid domain boundary conditions.
Otherwise, we interpolate the advected velocity of grid node $\ii$ at $\xx^{\text{upwind}}_{\ii}$ using the monotonic cubic spline as in \citet{fedkiw:2001:visual}.
Other scalar quantities, such as temperature and density, are advected using the same approach.

\subsubsection{Apply Forces} \label{section:apply_forces}

We apply a buoyancy force to the incompressible fluid.
%In equation TODO the fluid undergoes some body forces $\ff$.
Accordingly, we solve Equation \eqref{eq:body_forces}
after the advection step but before the projection step of the incompressible update, where $\ff$ accounts for buoyancy.
Specifically, after advection, we obtain an intermediate velocity $\uu^A$, which we update with body forces via
\begin{equation}
    \uu^* = \uu^A + \Delta t \ff. 
\end{equation}
The buoyancy force that we apply is
\begin{equation}
    \ff_\text{buoy} = \alpha (T - \bar{T}) \zz
\end{equation} 
as in \citet{nguyen2002physically}, where $\alpha > 0$, $T$ is the temperature, $\bar{T}$ is the ambient or room temperature, and $\zz$ is the unit vector pointing in the upward direction.
The projection step corrects $\uu^*$ to be divergence-free.

\subsubsection{Particle Advection} \label{section:particle_advection}

After enforcing incompressibility, we update smoke particle velocities and positions in a FLIP style:
\begin{equation}
    \vv_p^{n+1} = (1 - \alpha) \vv_p^{\text{PIC}} + \alpha \vv_p^{\text{FLIP}},\quad \xx_p^{n+1} = \xx_p^n + \dt \vv_p^n ,
\end{equation}
where $\vv_p^{\text{PIC}} = \sum_{\ii} N_{\ii} (\xx_p^n) \uu_{\ii}^{n+1}$ is interpolated from the updated fluid grid velocity and $\vv_p^{\text{FLIP}} = \vv_p^n + \sum_{\ii} N_{\ii} (\xx_p^n) \left(\uu_{\ii}^{n+1} - \uu_{\ii}^n\right)$ is from the change in grid velocity. Note that we adopted FLIP transfers because we found that the noise gave a better look for fire and smoke than the more stable APIC transfers. In fact, we use $\alpha = 0.99$ to get as much noise as we can without observing instability.

\subsection{Temperature Step} \label{section:temperature_step}

\subsubsection{Narrow-Band Style Temperature Advection} \label{section:temperature_p2g}

Similar to the advection scheme described in Section \ref{section:incompressible_advection}, we first transfer temperature from MPM particles and smoke particles to the temperature grid using linear APIC as in \citet{jiang:2015:apic} and \citet{chen:2021:momentum}: 
\begin{equation}
    m_{\ii}T_{\ii}^n = \sum_p m_p N_{\ii} (\xx_p^n) \left( T_p^n + (\xx_{\ii} - \xx_p^n) \cdot \nabla T_{p}^n \right) . \label{eq:temperature_P2G}
\end{equation}
For grid nodes that do not receive information from particles, we use the semi-Lagrangian advection scheme described in Section \ref{section:incompressible_advection} to update the grid temperature.
Note that temperature grid nodes are located at the cell center of the incompressible fluid grid. We linearly interpolate the fluid velocity to get the cell center velocity $\uu_{\ii}^c$. We choose to interpolate the velocity to the cell center rather than the temperature to the incompressible grid nodes to minimize unwanted diffusion. We summarize our scheme as
\begin{equation}
    T^A_{\ii} = 
    \begin{cases}
       T_{\ii}^n, & \text{for } m_{\ii} \neq 0, \\
        \text{semi-Lagrangian}(\xx_{\ii}, \uu_{\ii}^c, \dt), & \text{otherwise}.
    \end{cases}
\end{equation}

% okay now!
% \subsubsection{Particle Temperature Transfer to Particles (G2P)} \label{section:temperature_p2g}

% We transfer temperature from MPM particles and smoke particles to the temperature grid using linear APIC as in \citet{jiang:2015:apic} and \citet{chen:2021:momentum}:
% \begin{equation}
%     T_{\ii}^n = \sum_p m_p N_{\ii} (\xx_p^n) \left( T_p^n + (\xx_{\ii} - \xx_p^n) \cdot \nabla T_{p}^n \right) \label{eq:temperature_P2G}
% \end{equation}
% We use a linear transfer to preserve the sharpness of temperature.

\begin{figure}
    \centering
    \resizebox{0.45\textwidth}{!}{\includesvg{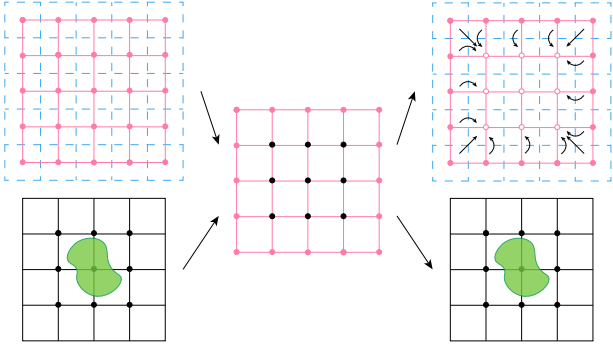}}
    \caption{After transferring the temperature from the MPM object (green) to the MPM grid, we overwrite the temperature grid nodes (pink), located at the cell centers of the incompressible grid (blue), with the MPM grid temperature from the MPM grid nodes that have positive mass (black). After updating the temperature on the temperature grid, we transfer the updated temperatures back to the MPM grid nodes with MPM mass, which we then transfer to the MPM particles. We then extrapolate the fluid temperature to the nodes on the temperature grid corresponding to the MPM grid nodes with positive mass (white/hollow pink).}
    \label{fig:temperature_step}
\end{figure}

% We transfer temperature from MPM particles and smoke particles to the temperature grid using
% \begin{equation}
%     T_{\ii}^n = \sum_p m_p N_{\ii} (\xx_p^n) \left( T_p^n + (\xx_{\ii} - \xx_p^n) \cdot \nabla T_{p}^n \right) \label{eq:temperature_P2G}
% \end{equation}
% We use a linear grid to preserve the sharpness of temperature (i.e. very hot object versus a cool ambient temperature).  TODO: Cite Jingyu's paper?

% \subsubsection{Merge MPM and Fluid Temperatures} \label{section:merge_temperatures}

% Before we can solve for the temperature in \eqref{diffusion_solve}, we need to merge the solid and fluid temperatures. We loop over all of the temperature grid nodes. If a grid node that has nonzero mass $m_{\ii}$ from Temperature P2G, we overwrite the fluid temperature at that grid node with the splatted MPM temperature. Otherwise, we use the fluid temperature at that grid node. See figure \ref{fig:temperature_step}.

\subsubsection{Temperature Solve} \label{section:temperature_solve}

After merging the fluid and MPM grid temperatures, we solve for the updated temperature. Equation \eqref{diffusion_solve} can be discretized in 3D as:
\begin{align} 
\rho c_p \dfrac{T_{i,j,k}^{n+1} - T_{i,j,k}^n}{\Delta t} & = \dfrac{K}{\Delta x^2} \Big( T_{i+1,j,k}^{n+1} + T_{i,j+1,k}^{n+1} + T_{i,j,k+1}^{n+1} - 6 T_{i,j,k}^{n+1} \\
& \quad \quad + T_{i-1,j,k}^{n+1} + T_{i,j-1,k}^{n+1} + T_{i,j,k-1}^{n+1} \Big) + H . \notag
\end{align}
We can isolate the time $t^{n+1}$ terms on one side, then solve for $T^{n+1}$ using the Hestenes-Stiefel formulation of the conjugate gradient method from \citet{golub2013matrix}.

Additionally, we can apply any temperature boundary conditions at this step.

\subsubsection{Temperature Extrapolation} \label{section:temperature_extrapolation}

Fluid temperatures located where the solid is located may be unreliable for advection.
We extrapolate fluid temperature values from the well-defined fluid temperatures to the rest of the grid using a breadth-first search algorithm as described in \citet{bridson2015fluid}. See Figure \ref{fig:temperature_step}.

\subsubsection{Grid Temperature Transfer to Particles (G2P)} \label{section:temperature_g2p}

We again use linear APIC to transfer temperature and temperature gradient from the temperature grid to particles using the following updates:
\begin{equation}
    T_p^{n+1} = \sum_{\ii} \hat{T}_{\ii}^{n+1} N(\xx_{\ii}), \quad
    \nabla T_{p}^{n+1} = \sum_{\ii} \hat{T}_{\ii}^{n+1} \nabla N (\xx_{\ii}). 
    \label{eq:temperature_G2P}
\end{equation}

\subsection{Ignition Step} \label{section:ignition_step}

Every MPM particle can be in one of four states: original (O), about to burn (TB), burning (B), and burnt (D). We use $s_p$ to denote the current state of particle $\xx_p^n$.
To start ignition, we choose at least one particle and set it as burning ($s_p = \text{B}$) at the beginning of each simulation.
If an MPM particle is burning, we update its fuel level and temperature, and ignite neighboring particles as described in the substeps below. 
We also update the fuel level and temperature as described in Sections \ref{section:update_fuel_level} and \ref{section:update_temperature} for each smoke particle.  %If particle is set to burn, we count down the time to ignition. If the time to ignition has elapsed then we set status to burning. If a particle is not burning or burned then we skip the following substeps and continue to the next time step. 

\subsubsection{Update Fuel Level} \label{section:update_fuel_level} 

Each particle is assigned an initial fuel $F_0$. For all particles with $s_p = \text{B}$, we decrease the fuel according to the differential equation
\begin{equation}
     \dfrac{dF_p}{dt} = -\gamma F_p , \label{fuel_update}
\end{equation}
where $\gamma$ is some positive parameter. The solution to Equation \eqref{fuel_update} is
\begin{equation}
 F_p^n = F_0 e^{-\gamma t},
\end{equation}
which we can use directly in our ignition step. Notice that this ensures the fuel will always be positive. When $F_p^n$ is less than the fuel threshold $F_{\min}$, we set the particle state to burnt ($s_p = \text{D}$). 

\subsubsection{Update Temperature} \label{section:update_temperature}

Burning particles ($s_p = \text{B}$) increase in temperature based on their fuel level. We model this as
\begin{equation}
    \dfrac{dT_p}{dt} = \beta F_p^n \dt\label{temperature_ignition_update}
\end{equation}
where $\beta$ is some positive parameter. The update to Equation \eqref{temperature_ignition_update} can be written as
\begin{equation}
T_p^{n+1} = T_p^n + \beta F_p^n \dt.
\end{equation}
We then cap the temperature update to a maximum allowed temperature for the object, $T_\text{max}$.

%\subsubsection{Determine particle burning state} \label{section:determine_burning_state}

%When a burning particle's fuel level is below a fuel level threshold, we classify the particle as burned. 

\subsubsection{Ignite neighboring particles} \label{section:ignite_neighboring_particles}

For every burning particle $\xx_p^n$, we find its closest neighboring particle $\xx_q^n \in \partial \mathcal{S}$, where $\partial \mathcal{S}$ is the surface of $\mathcal{S} \coloneq \{\xx_r^n | s_r = \text{O}\}$. If $T_q^n$ is above the ignition threshold $T_{\text{ignition}}$, then we set $s_q = \text{TB}$. To control the speed of the flame front across the surface of the object, we set an ignition delay by calculating a time to burn using the equation
\begin{equation}
t_{\text{delay}} = \dfrac{||\xx_p^n - \xx_q^n||_2}{c_{\text{flame}}} ,
\end{equation}
where $||\xx_p^n - \xx_q^n||_2$ is the distance to the neighboring particle and $c_{\text{flame}}$ is the desired flame front propagation speed.
When this time to burn has elapsed, we set $s_q = \text{B}$.

\subsection{Acceleration Structure}
Throughout our simulation loop, when dealing with particles, we use a spatial hashing technique to accelerate the search for nearby particles.
Particles are binned using a spatial hash grid of size $\dx$.

In Section \ref{section:resample_smoke}, to find the closest boundary particle $\bb^n$ to a burning MPM particle $\xx_p^n$, we first build a hash table for all boundary particles and compute the hash grid locations of particle $\xx_p^n$.
Since the particle level set is constructed using a particle radius of $\frac{\sqrt{d}}{2}\dx$, $\bb^n$ is guaranteed to be found within a 1-grid-cell radius of $\xx_p^n$, i.e., 8 adjacent cells in 2D or 26 adjacent cells in 3D.

In Section \ref{section:ignite_neighboring_particles}, we follow a similar approach to find the next particle to ignite. We first spatially hash all MPM particles with $s_p = \text{O}$. Then, for each boundary particle $\bb^n$, we find its closest MPM particle $\xx_p \in \mathcal{S}$ and add $\xx_p$ to set $\partial \mathcal{S}$. We construct a hash table for particles in $\partial \mathcal{S}$ and use it to find the closest particle $\xx_q^n \in \partial \mathcal{S}$ to particle $\xx_p^n$.

\subsubsection{Sparse Implementation of MPM}
% \textcolor{red}{Sparse MPM section somewhere?...}
For each MPM or smoke particle, the particle information will only be transferred to the nearby $2^d$ or $3^d$ grid nodes for linear or quadratic interpolation functions respectively in $d$ dimensions. 
Particles only activate a small portion of grid nodes in a rectangular computational domain.
In our implementation, we only store the information on the activated grid nodes to diminish unnecessary storage.

\subsection{Mesh-Based Fracture}
\label{sec:fracture}

When a solid object burns and undergoes shrinking, we expect deformations in its surface.
We utilize a post-process, mesh-based fracture scheme as in \citet{wang:2019:fracture} to model this deformation.
This method identifies any broken edges that occur throughout the simulation and introduces duplicated nodes with smoothing.
We use this in our log example (Section \ref{sec:log}).

\section{Examples}
\label{sec:examples}

%We demonstrate our method on a two-dimensional test case and on a series of three-dimensional examples.
%In the examples, the temperature boundary conditions are Dirichlet boundary conditions with room temperature 298K. 

We demonstrate our method on a two-dimensional test case and on a series of three-dimensional examples. 
All examples were run with $K_{\text{air}} = 0.01$, $K_{\text{solid}} = 0.1$, $c_{p, \text{air}} = 1$, $c_{p,\text{solid}} = 1$, $\bar{T} = 298$ K. The three-dimensional examples used $T_{\text{ignition}} = 600$ K, an approximate temperature for wood combustion as seen in \citet{babrauskas:2002:ignition,yudong:1992:measurement}.
Runtime performance is listed in Table \ref{tab:performance}, and additional parameters used are listed in Table \ref{tab:parameters}. 
Simulations were run on an AMD workstation equipped with 128 threads. 
$\dt$ is chosen in an adaptive manner restricted by a CFL condition such that no particles are allowed to travel more than a portion of $\dx$ in each time step.

\begin{table}
    \centering
    \begin{tabular}{lccccr}
        \hline 
        Simulation & $d$ & $\dx$ & Time & Elements & Particles \\
        \hline
        Squares & 2 & 1/16 & 0.537 & $-$ & 1k \\
        Log & 3 & 1/64 & 71.1 & 73.6k & 60.3k \\
        Incense & 3 & 1/128 & 701 & $-$ & 1.62M  \\
        Match & 3 & 1/256  & 1180 & $-$ & 124k \\ \hline 
    \end{tabular}
    \caption{All simulations were run on an AMD workstation with 128 threads. Simulation time is measured in average seconds per frame. Elements denotes number of tetrahedra in the volumetric mesh, where applicable. Particles denotes the number of MPM particles in the simulation.}
    \label{tab:performance}
\end{table}

\begin{table}
    \centering
    \resizebox{\columnwidth}{!}{%
    \begin{tabular}{lcccc}
        \hline 
        Parameters & Squares & Log & Incense & Match \\
        \hline 
         %$K_{\text{air}}$ & 1e-2 & 1e-2 & 1e-2 & 1e-2 \\
         %$K_{\text{solid}}$ & 0.1 & 0.1 & 0.1  &  0.1 \\
         %$c_{p, \text{air}}$ & 1 & 1 & 1 & 1 \\
         %$c_{p, \text{solid}}$ & 1.7 & 1.7 & 1.7 & 1.7 \\
         %$\bar{T}$ & 298 K & 298 K & 298 K & 298 K \\ 
         %$T_{\text{ignition}}$ & $-$ & 600 K & 600 K & 600 K \\
         Buoyancy $\alpha$ & $-$ & 1e-1 & 1e-2 & 1e-2 \\
         Shrinking Coefficient $c_{\text{shrink}}$ & $-$ & 1.00001, 1.00005 & $-$ & 1.01 \\
         Initial Fuel $F_0$ & 1 & 1 & $F_0 \in [1,2]$ & 1 \\    % initial fuel
         Fuel Level Threshold $F_{\min}$ & 0.3 & 0.3 & 0.3 & 0.3 \\ % fuel threshold
         Fuel Level Coefficient $\gamma$ & 1, 10 & 1e-3 & 1 & 10 \\ % fuel coefficient
         Temperature Coefficient $\beta$ & $-$ & 1e10 & 1e10 & 1e3 \\ % temperature increase coefficient
         Flame Front Speed $c_{\text{flame}}$ & 0.03, 0.1 & 0.08 & 0.01 & 0.015 \\ 
         \hline 
    \end{tabular}%
    }
    \caption{Main parameters used in the simulations in Section \ref{sec:examples}. }
    \label{tab:parameters}
\end{table}

\subsection{2D Burning Squares}

\begin{figure}[!ht]
    \centering
    \includegraphics[width=0.65\columnwidth]{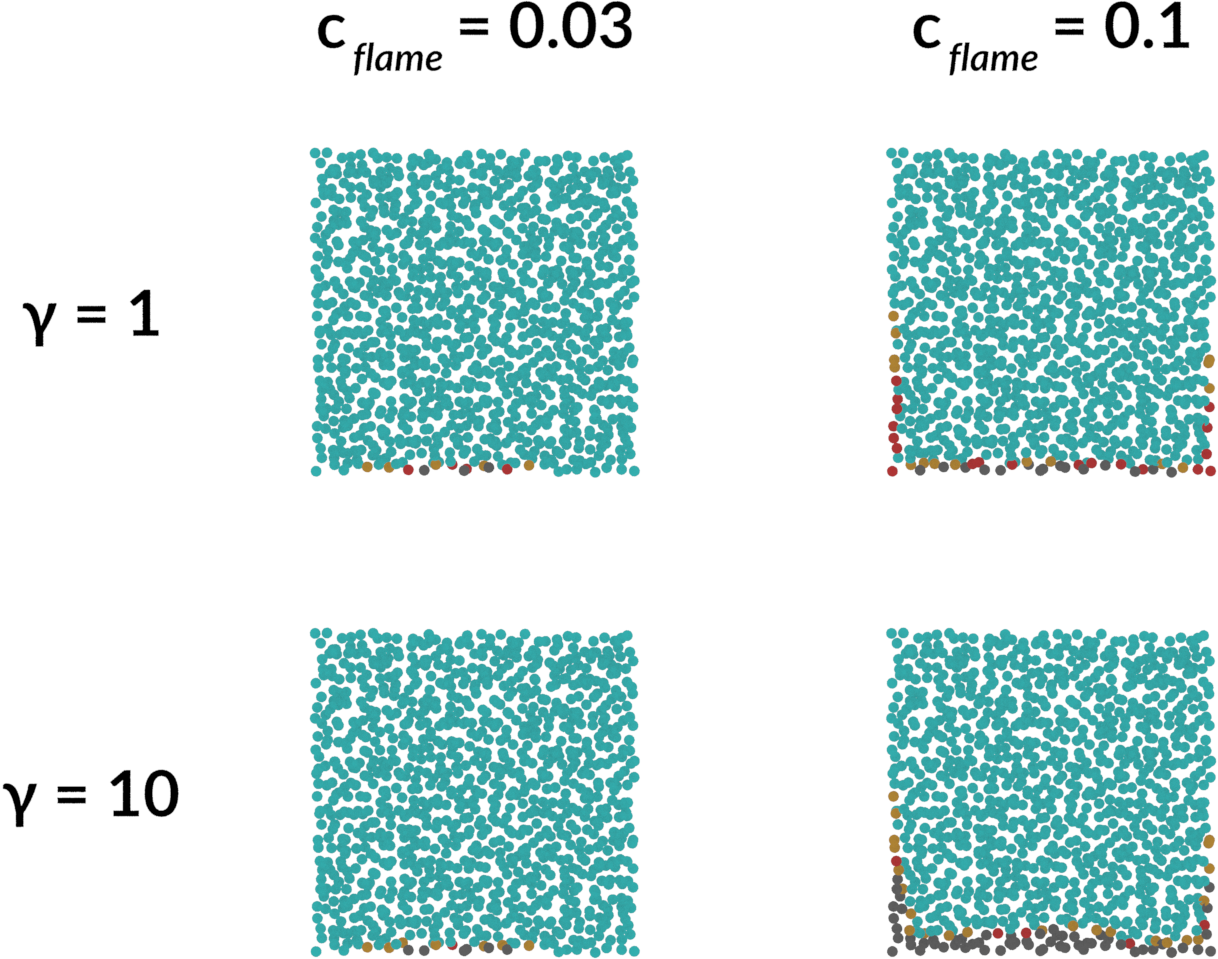}\vspace{2em}
    \includegraphics[width=0.65\columnwidth]{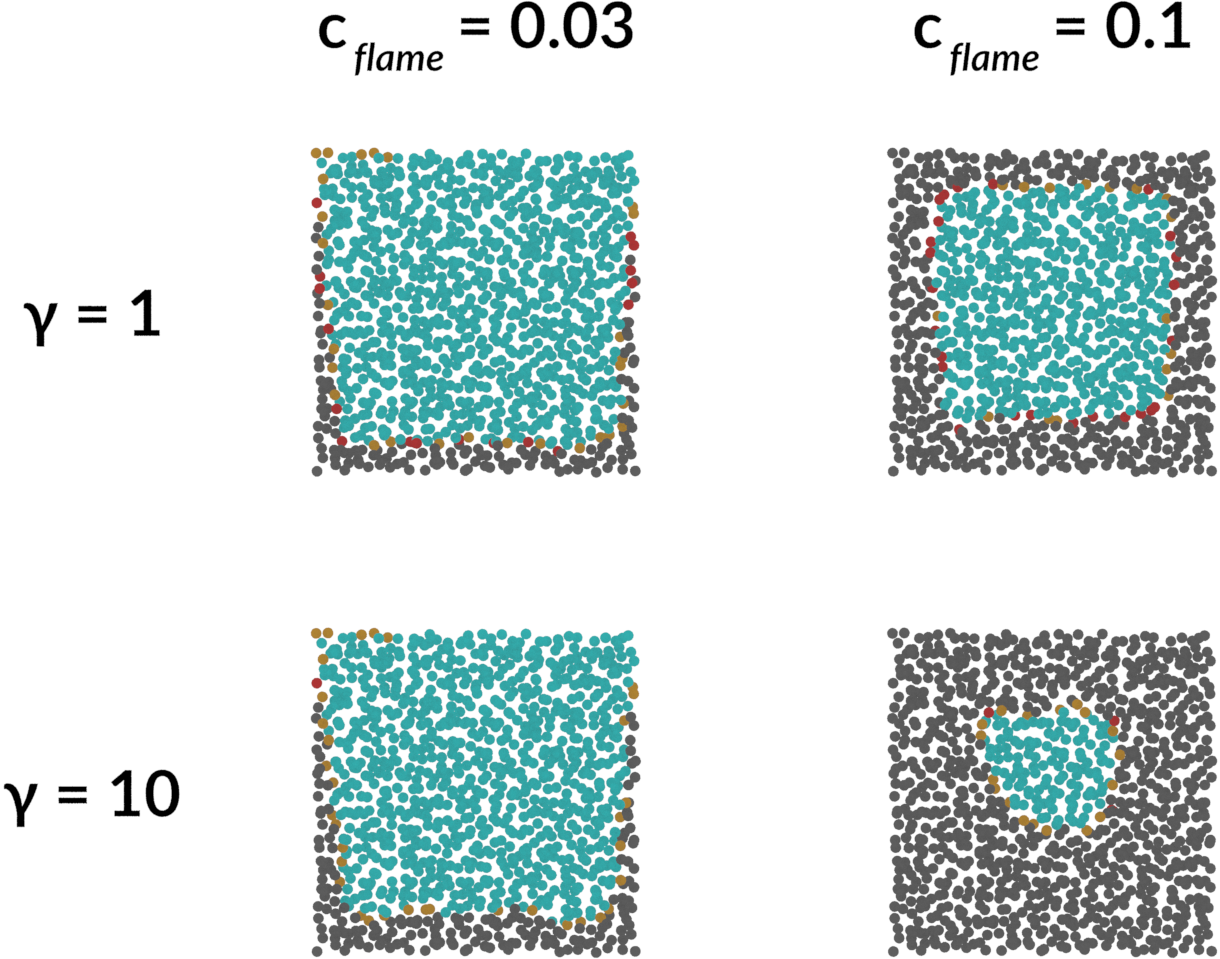}
    \caption{Squares are ignited at the center of the bottom edge of the domain. From the burning particles (red), we determine their neighboring particles on the surface that are hot enough and mark their status as about to burn (orange) after the time to ignite (determined by $c_\text{flame}$) is satisfied. At each step, we update a particle's fuel level according to the fuel coefficient $\gamma$, and once this fuel level has fallen below a fuel threshold, we mark the particle as burnt (gray). The burning status of particles from frames 100 and 600 are shown above and below, respectively.}
    \label{fig:burning_squares}
\end{figure}

To illustrate the directability of our method, we demonstrate the fuel update and ignition of neighboring particles in our ignition step on several squares with different fuel coefficients and flame propagation speed coefficients.
Results are shown in Figure \ref{fig:burning_squares}.
Intuitively, the larger the flame propagation coefficient, the faster burning will spread throughout the object, and the larger the fuel coefficient, the quicker a particle will burn.
Other parameters, such as diffusion coefficients $K$, temperature increase parameter $\beta$, initial fuel level per particle $F_0$, and fuel threshold $F_{min}$, will also affect the burning propagation.

\subsection{3D Log}
\label{sec:log}

\begin{figure}[!ht]
    \centering
    \begin{subfigure}[h]{0.49\columnwidth}
         \centering
         \includegraphics[width=\columnwidth]{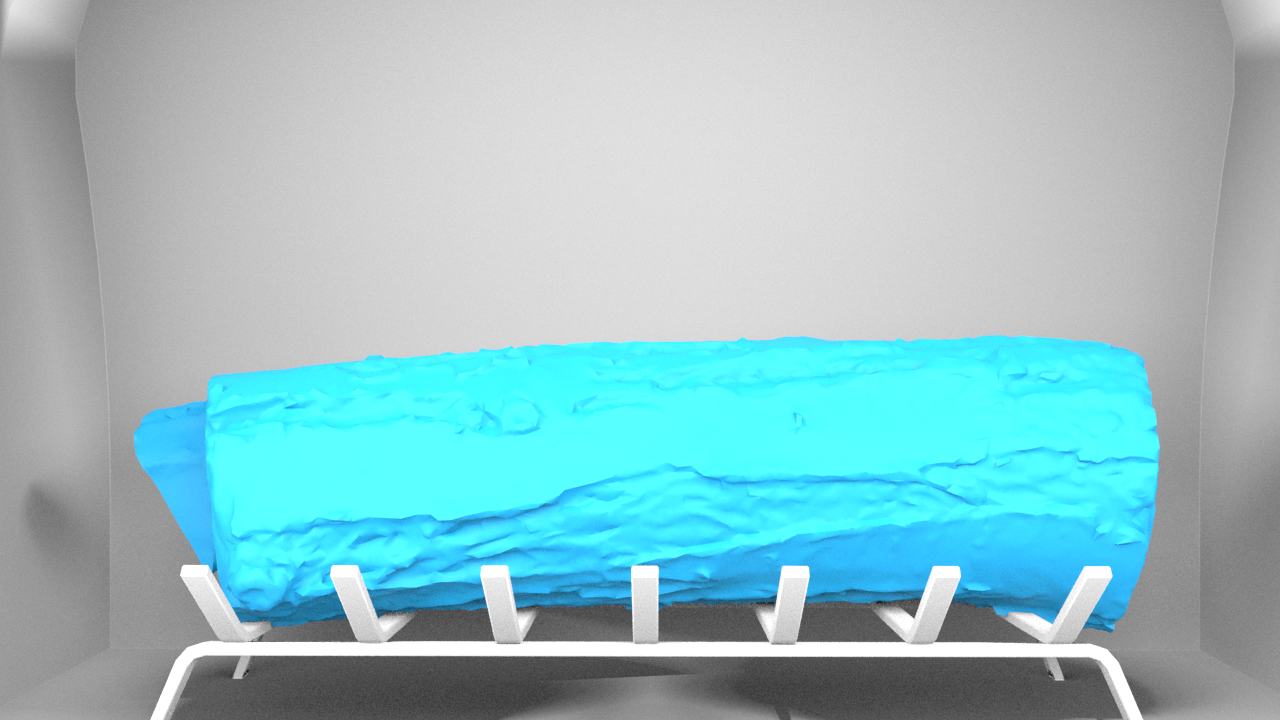} 
         \caption{Frame 0\vspace{0.5em}}
    \end{subfigure}
    \begin{subfigure}[h]{0.49\columnwidth}
         \centering
         \includegraphics[width=\columnwidth]{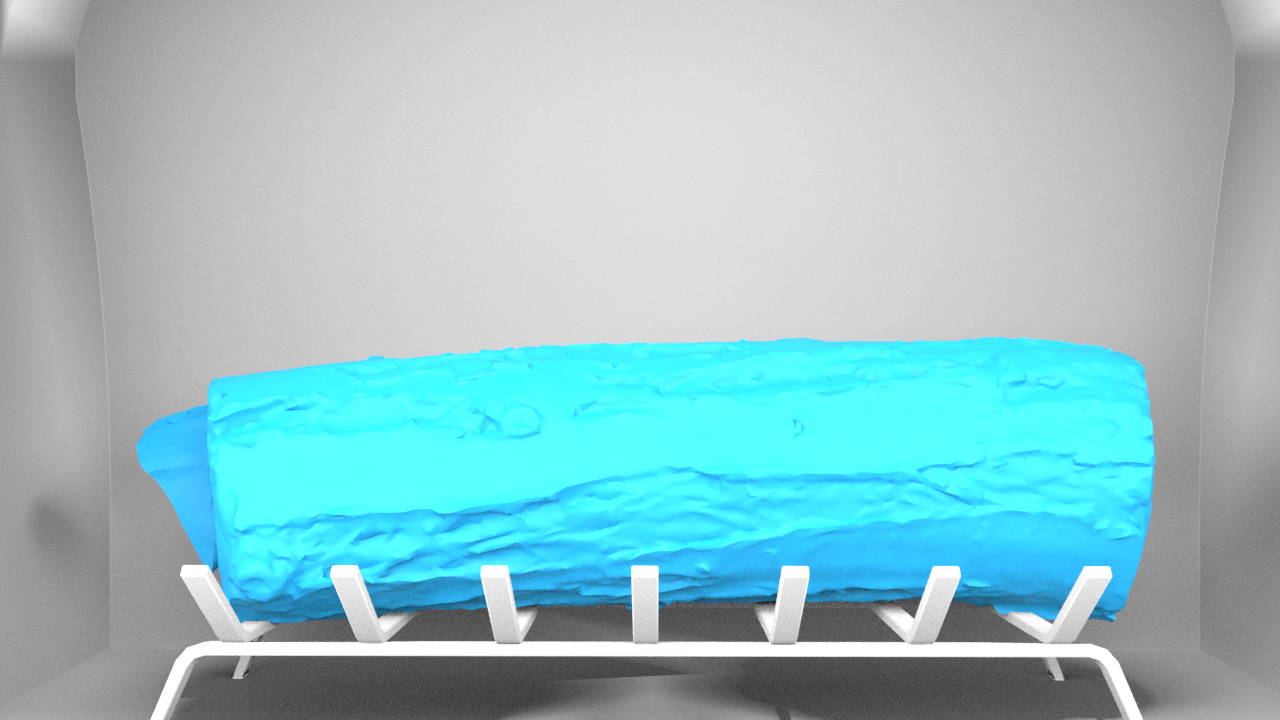}
         \caption{Frame 300\vspace{0.5em}}
    \end{subfigure}
    \begin{subfigure}[h]{0.49\columnwidth}
         \centering
         \includegraphics[width=\columnwidth]{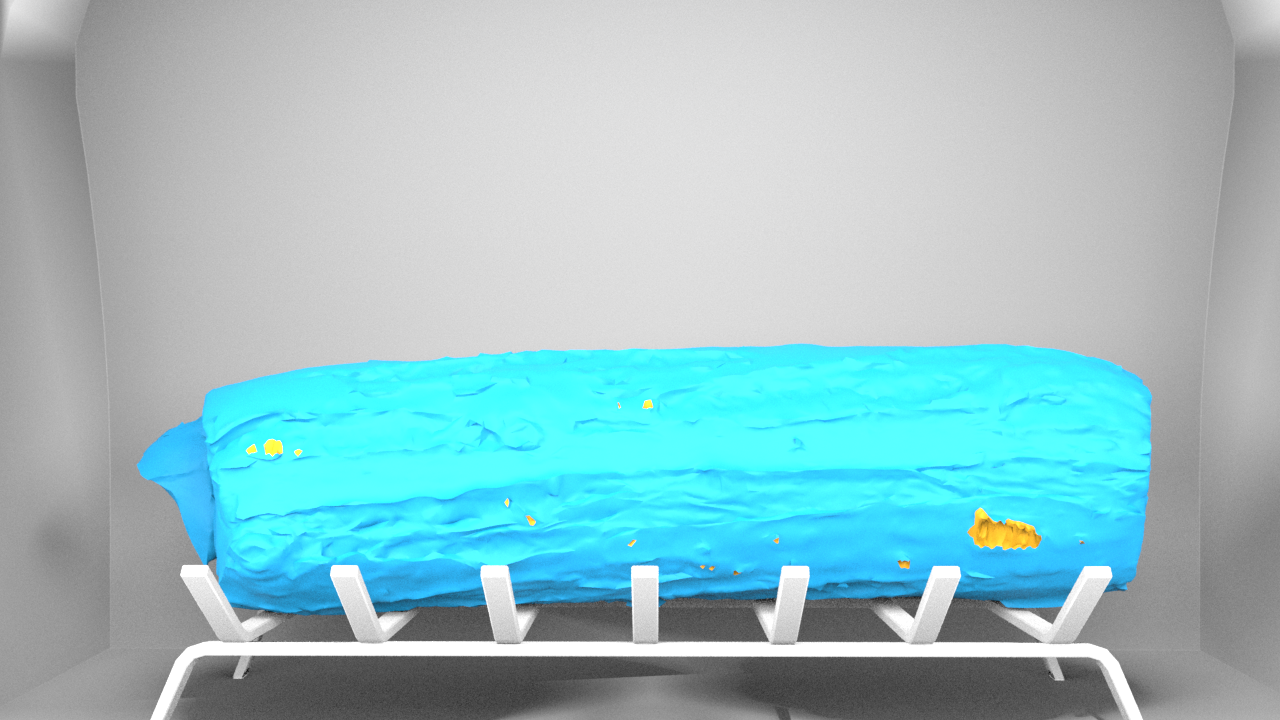}
         \caption{Frame 500\vspace{0.5em}}
    \end{subfigure}
    \begin{subfigure}[h]{0.49\columnwidth}
         \centering
         \includegraphics[width=\columnwidth]{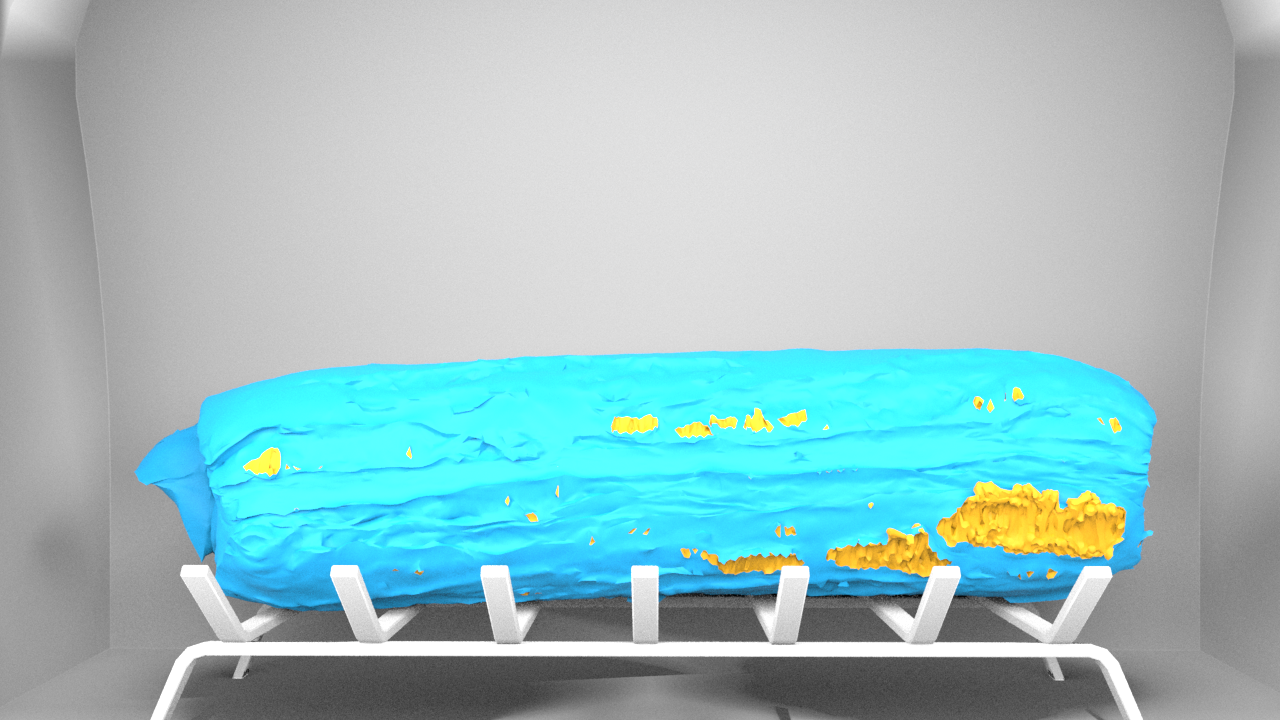}
         \caption{Frame 600\vspace{0.5em}}
    \end{subfigure}
    \caption{Views of the log example focusing on the shrinking and fracturing of the log.  Yellow indicates internal faces of the log mesh that become exposed due to mesh-based fracture as the log shrinks and deforms.}
    \label{fig:debug_log}
\end{figure}

\begin{figure}[!ht]
    \centering
    \includegraphics[width=\columnwidth]{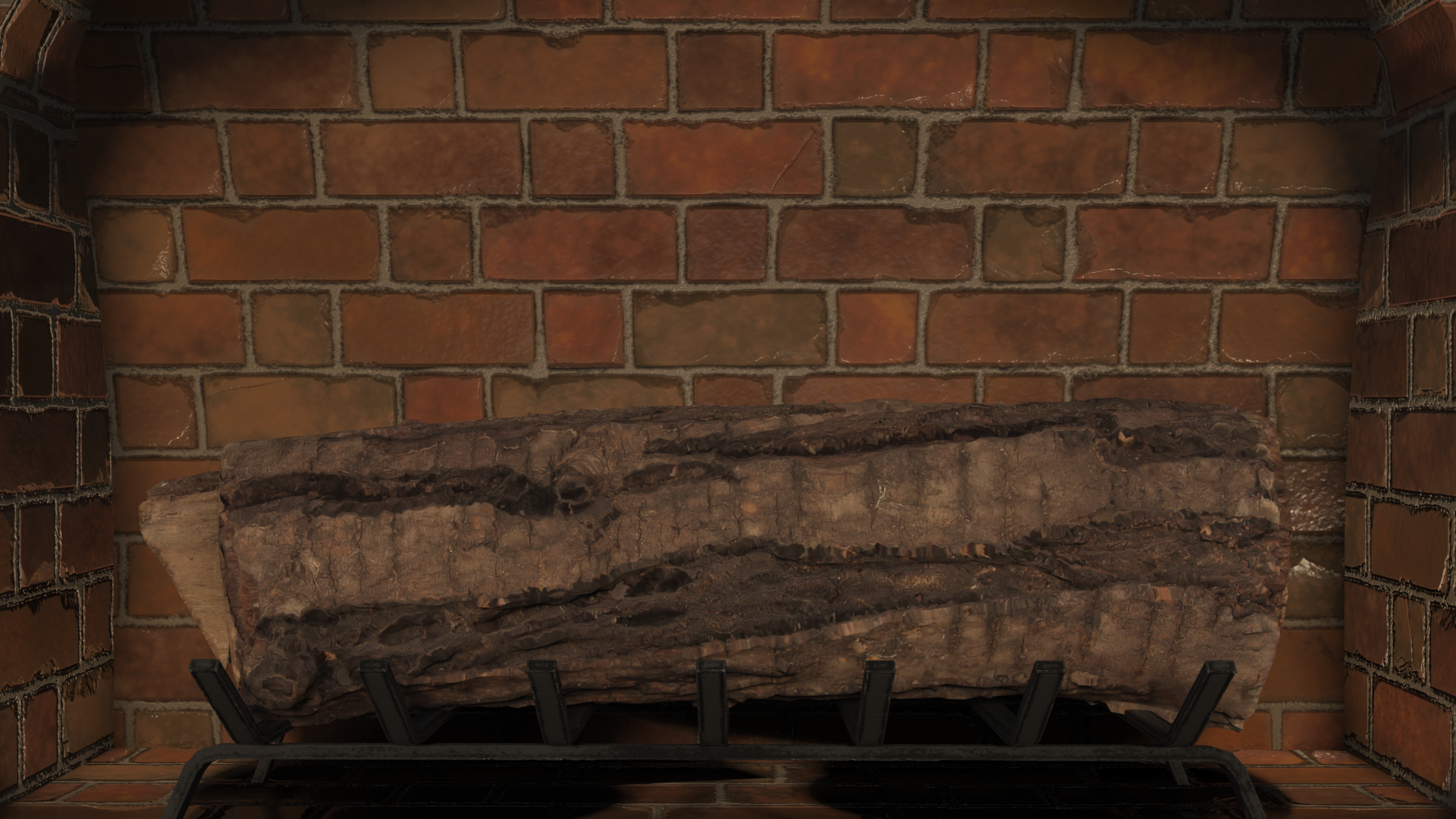}
    \includegraphics[width=\columnwidth]{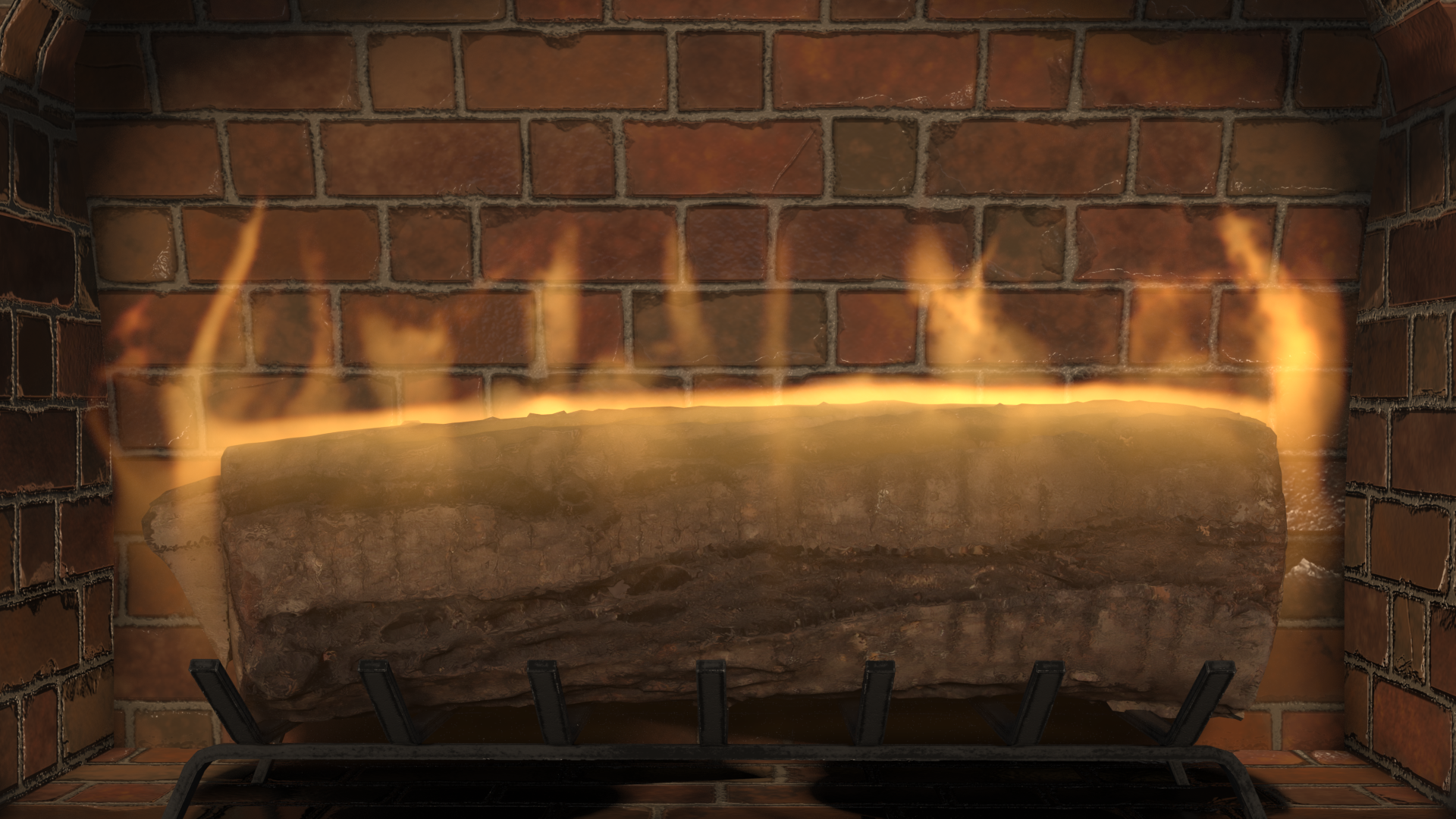}
    \includegraphics[width=\columnwidth]{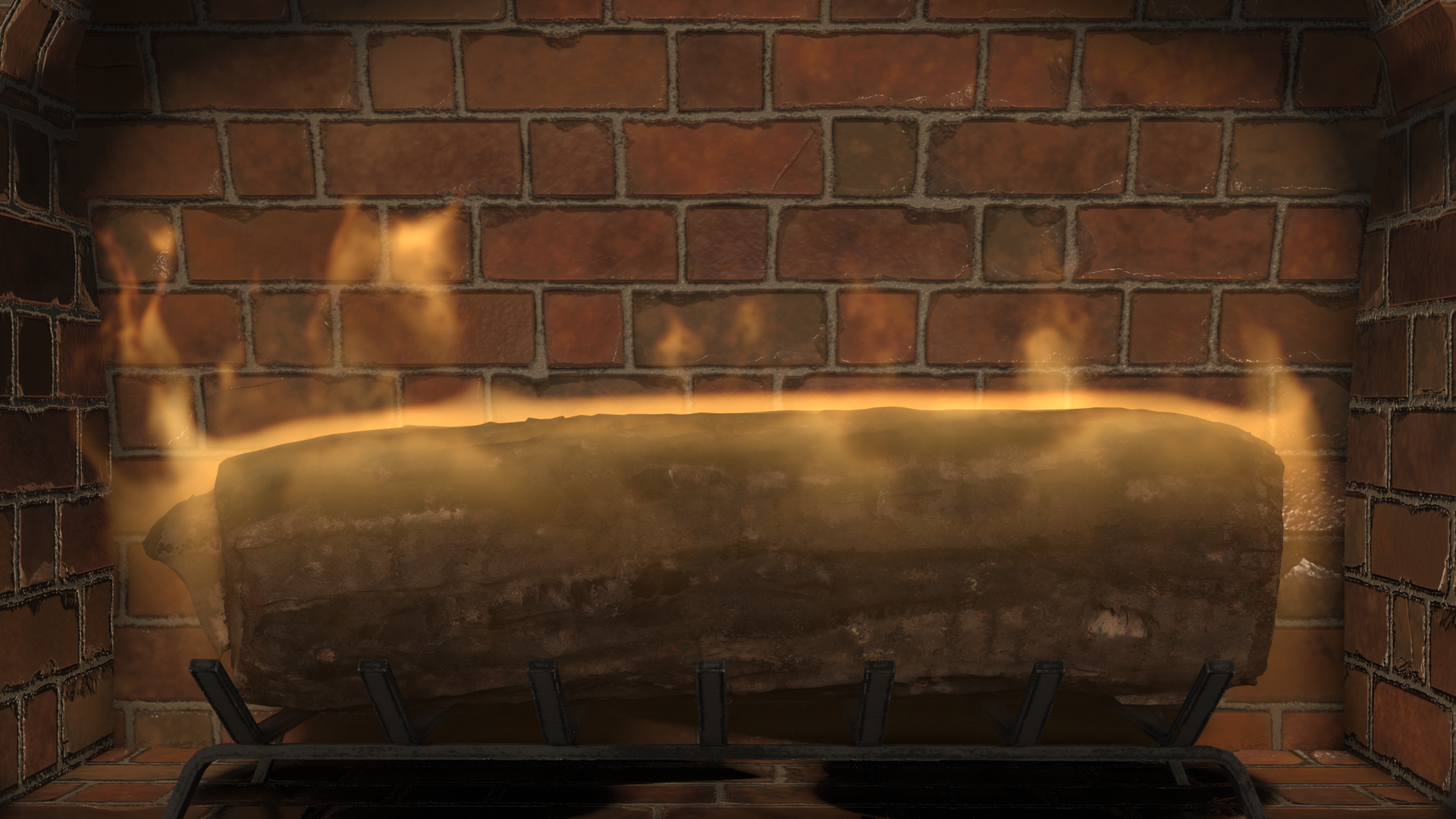}
    \includegraphics[width=\columnwidth]{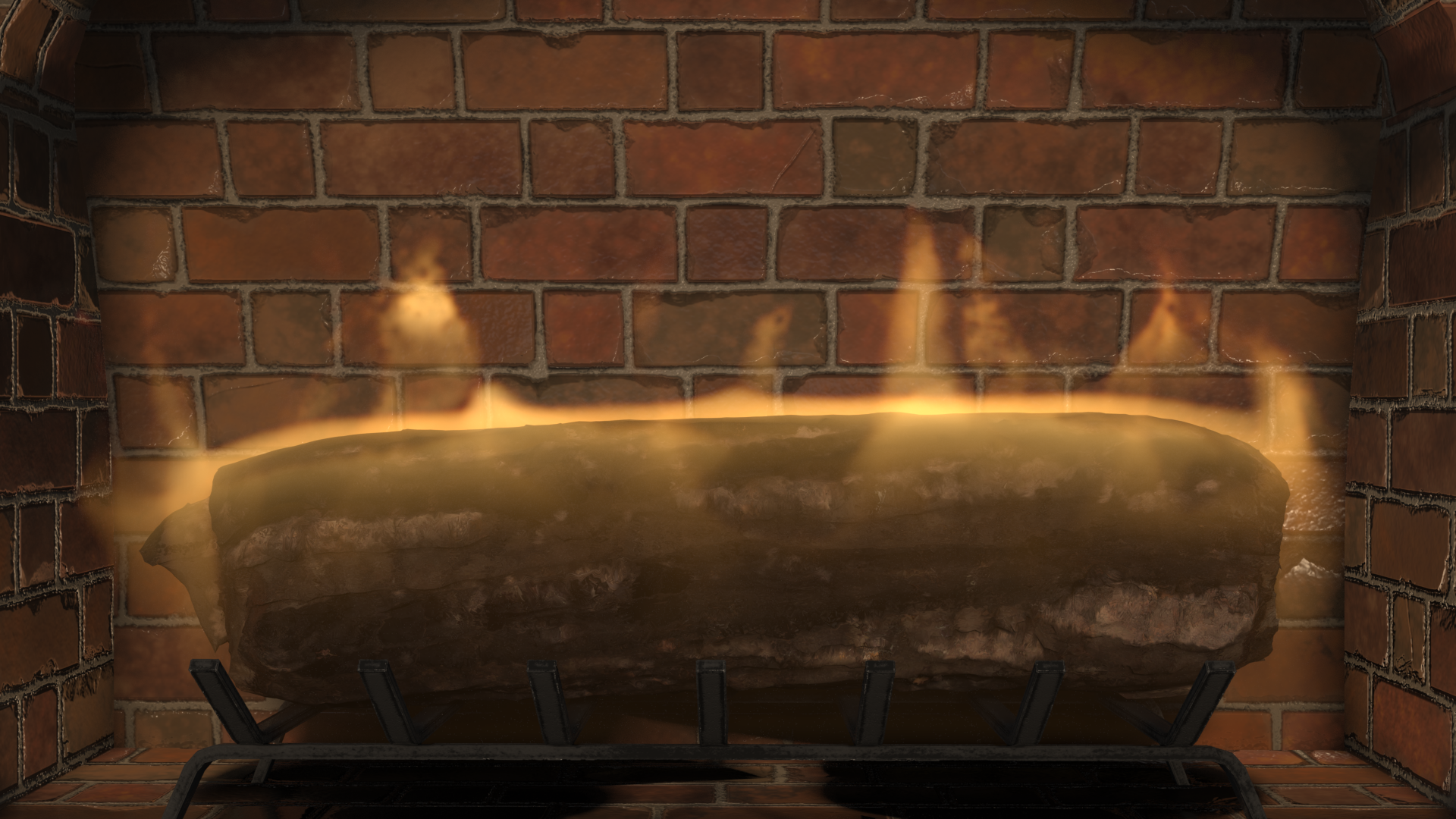}
    \caption{A log undergoes anisotropic shrinking and cracks as it burns. The mesh deforms, which we model with a post-process fracturing method. Frames 0, 300, 500, and 600 are shown.}
    \label{fig:burning_log}
\end{figure}

Figure \ref{fig:burning_log} illustrates a three-dimensional example where a log is ignited and the flame front spreads along the surface of the log.
The log undergoes anisotropic shrinking in both the radial and longitudinal directions using a radial shrinking coefficient $c_{\text{shrink}} = 1.00005$ and a longitudinal shrinking coefficient $c_{\text{shrink}} = 1.00001$.
This shrinking causes the mesh to deform, which we model with a post-process fracturing method (Section \ref{sec:fracture}).
%We show this simulation in two different ways.
%First, we show the log burning in a fireplace scene.
In the scene of Figure \ref{fig:burning_log}, we darken the texture of the log based on the fuel level of the MPM particles.
We also show a view of the log by itself and color the mesh based on its fractured surface in Figure \ref{fig:debug_log}.
Blue denotes the original surface of the log, and yellow denotes the fracturing.
We remark that even with relatively few computational resources (this simulation uses a resolution of $64$ cells per grid direction), we can achieve visually plausible results for a combusting solid.
%See Figure \ref{fig:burning_log}. TODO: Low res simulation, Even with relatively few computational resources, we can achieve visually plausible results

\subsection{3D Incense}
\label{sec:incense}

\begin{figure}[!ht]
    \centering
    \includegraphics[width=1.0\columnwidth]{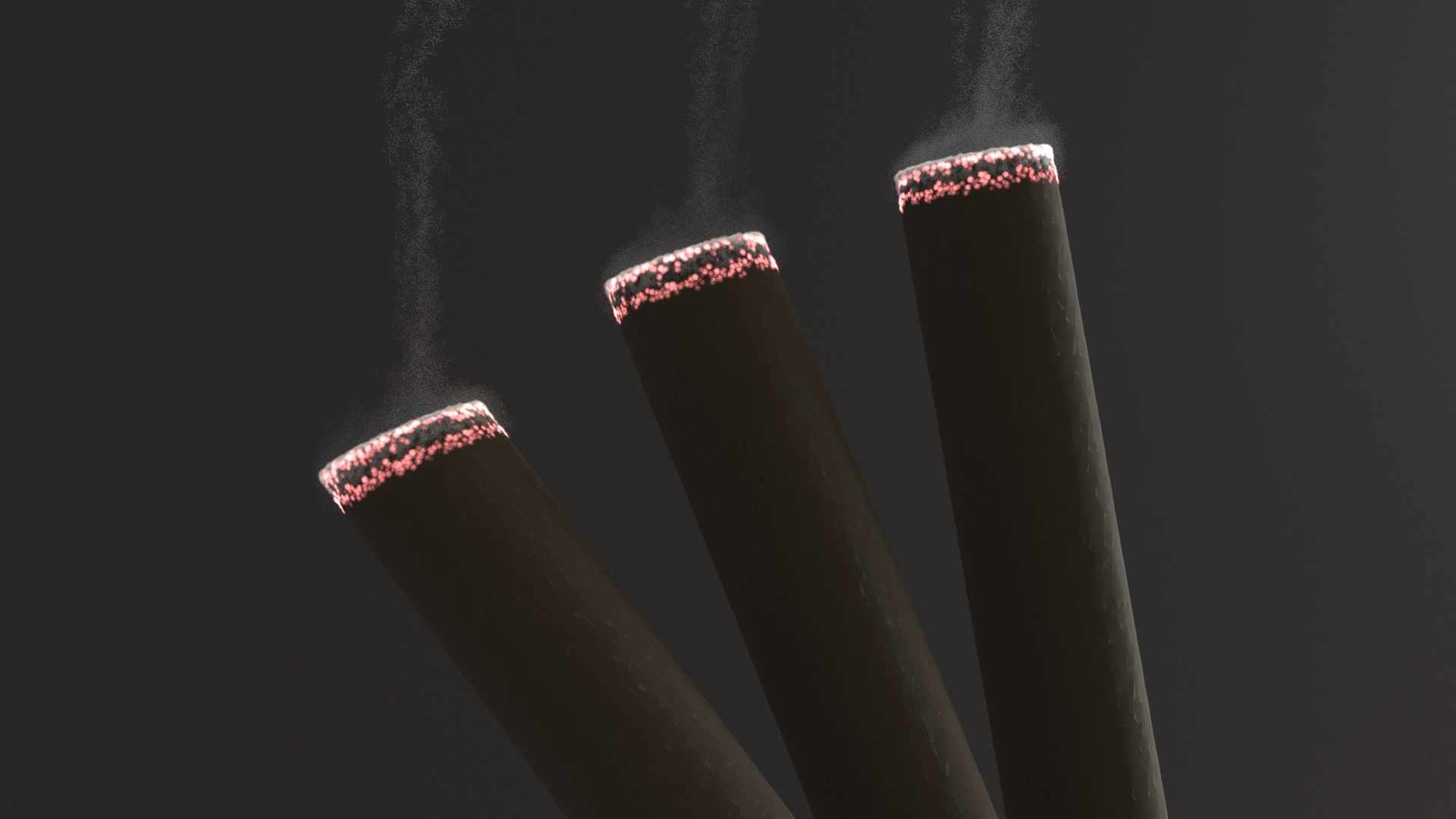}
    \includegraphics[width=1.0\columnwidth]{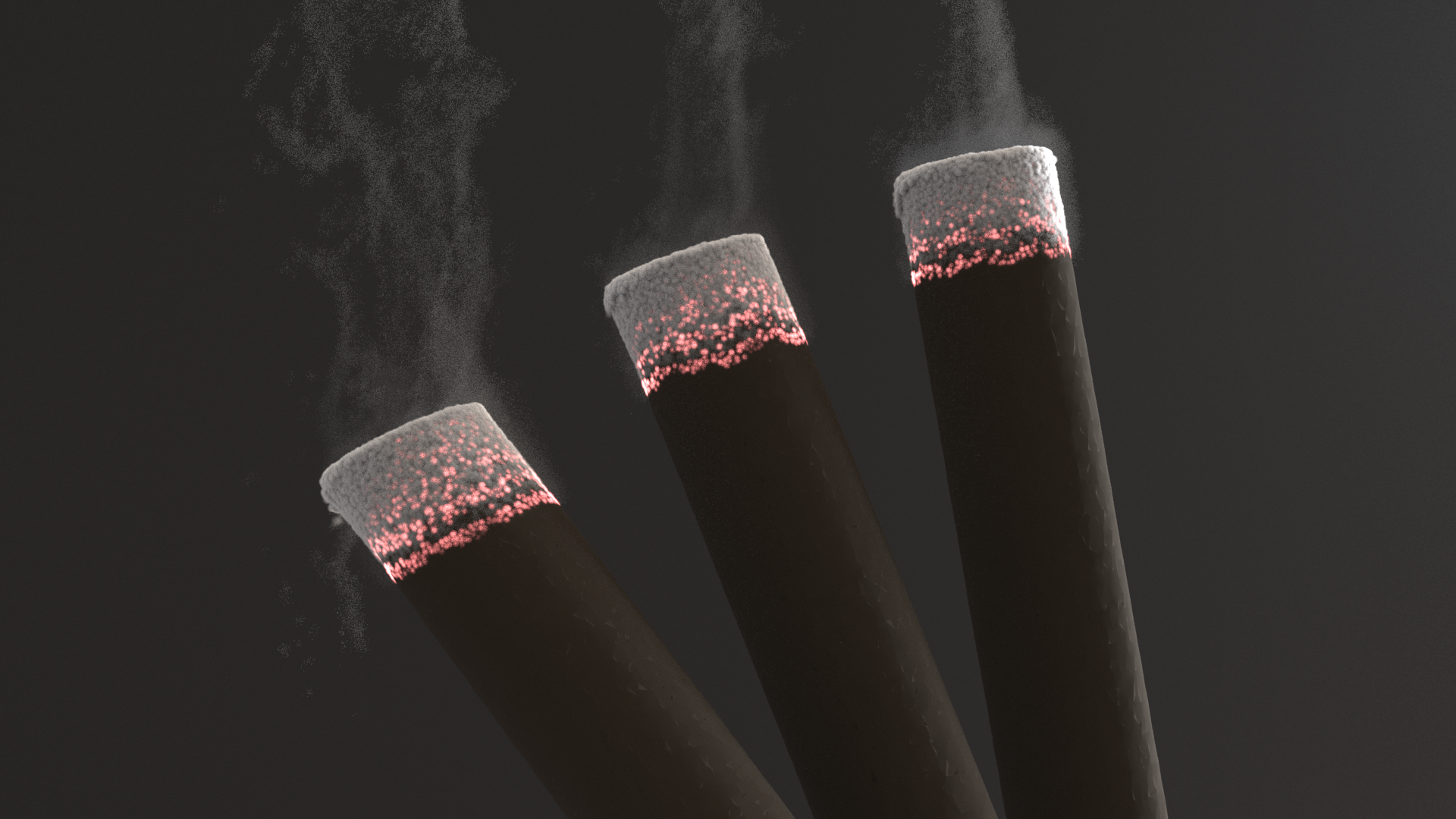}
    \includegraphics[width=1.0\columnwidth]{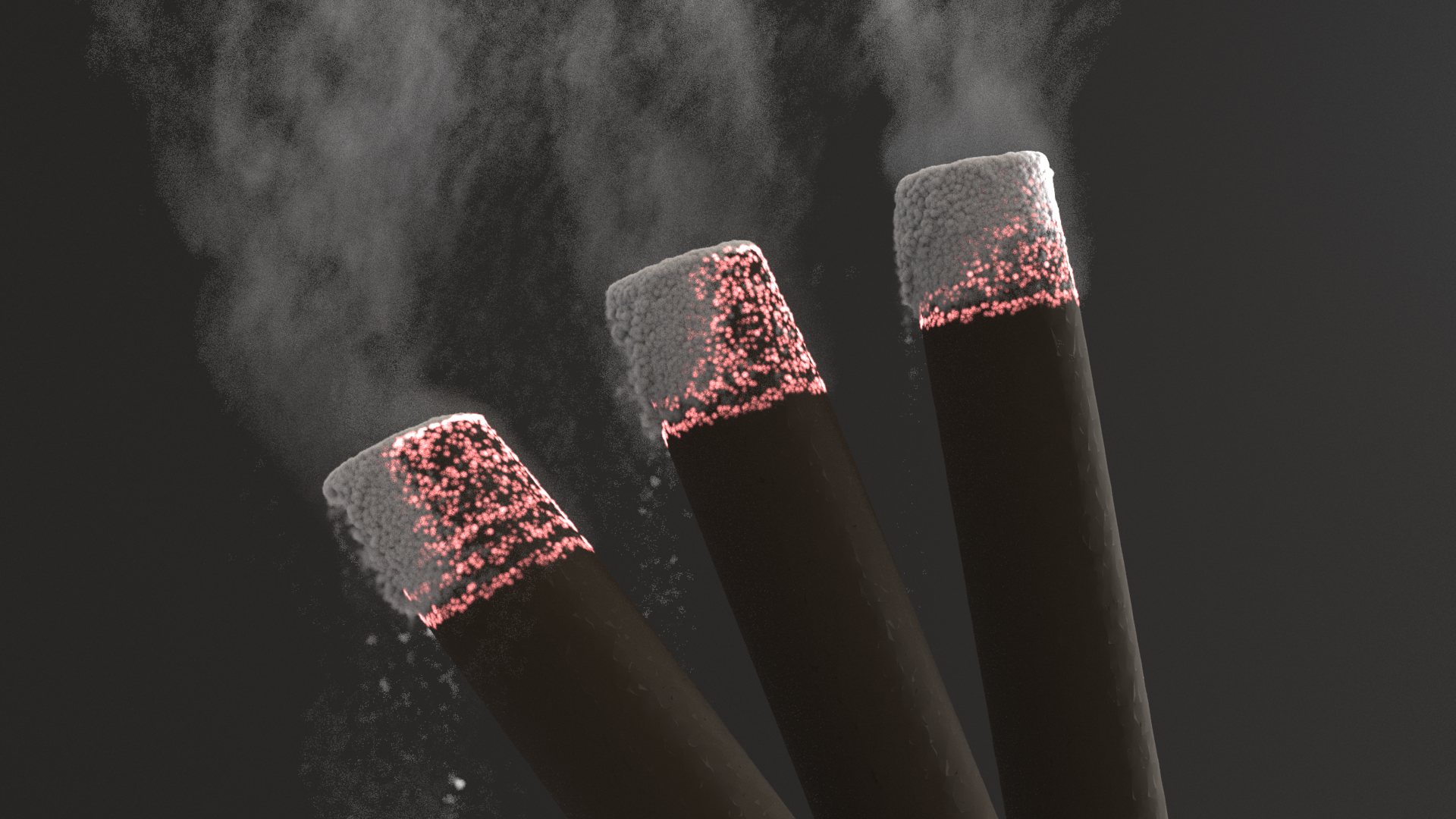}
    \caption{Burning incense undergoes a constitutive model change as it combusts. Our method is able to simulate the falling ash of the incense.}
    \label{fig:burning_incense}
\end{figure}

We next show a simulation of burning incense sticks, which better highlights the roles of smoke particles and constitutive model change in our framework.
Selected frames of this simulation are seen in Figure \ref{fig:burning_incense}.
We initialized the incense sticks to have a radially based initial fuel level between 1 and 2 to prevent the particles near the surface of the incense from burning too quickly. 
When the fuel capacity of a burning particle reaches a fuel level threshold, the particle is set as burnt and undergoes constitutive model change from an elastic to elastoplastic material (Section \ref{sec:update_phases}).
This constitutive model change causes the burnt particles to drop with gravity as the incense burns.
We use the burning status $s_p$ to model the different textures of the incense. Additionally, we model the emission of the burning particles based on their fuel level.

\subsection{3D Match}
\label{sec:match}

\begin{figure}[!ht]
    \centering
    \begin{subfigure}[h]{\columnwidth}
         \centering
         \includegraphics[width=0.63\columnwidth]{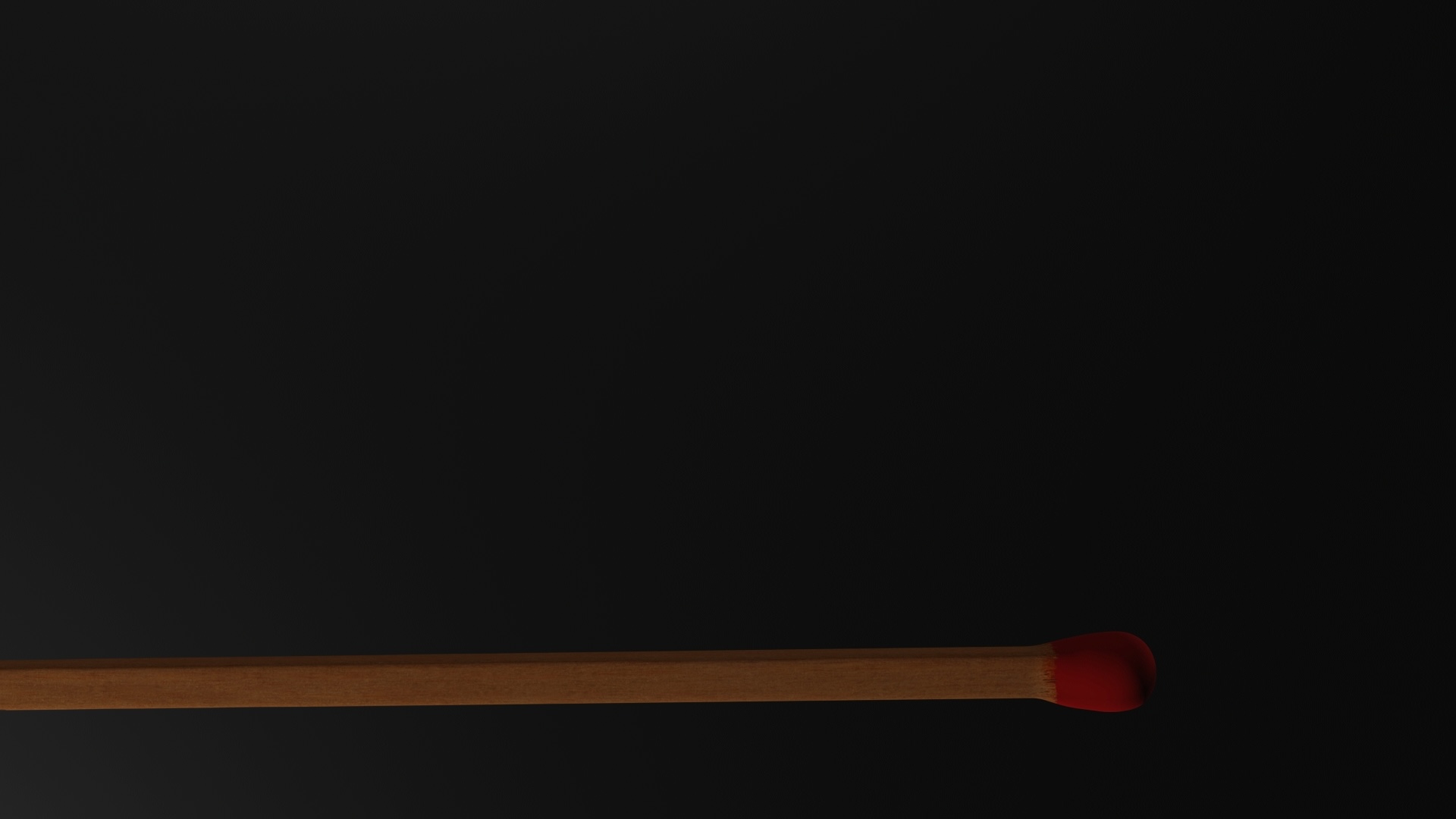} \includegraphics[width=0.355\columnwidth]{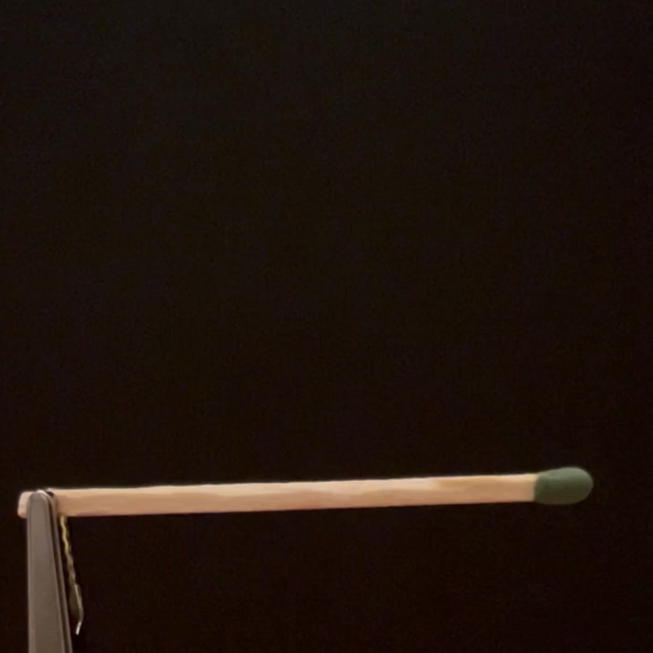}
         \caption{Left: Frame 1 \vspace{0.5em}}
    \end{subfigure}
    \vspace{0.5em}
    \begin{subfigure}[h]{\columnwidth}
         \centering
         \includegraphics[width=0.63\columnwidth]{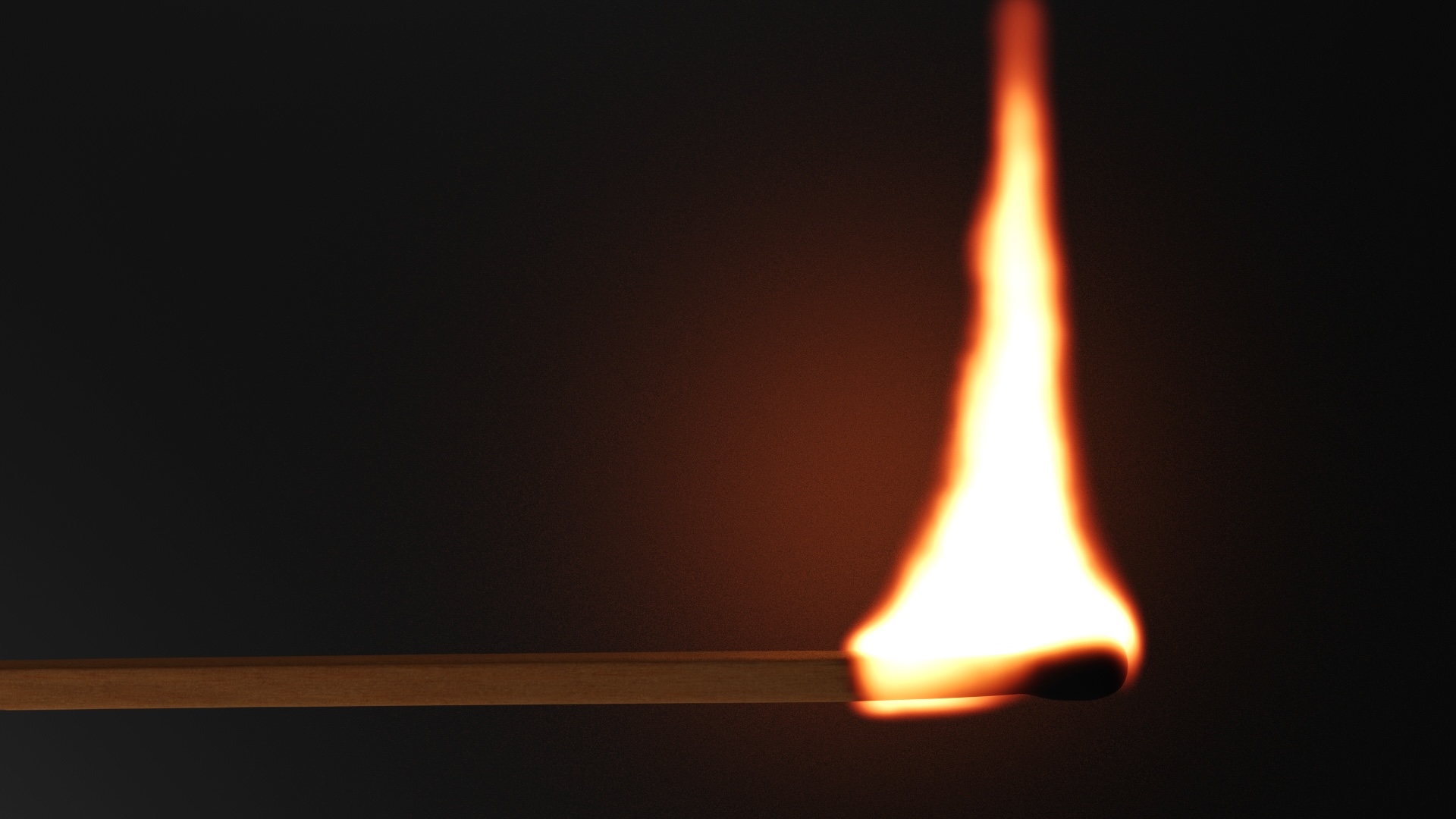} \includegraphics[width=0.355\columnwidth]{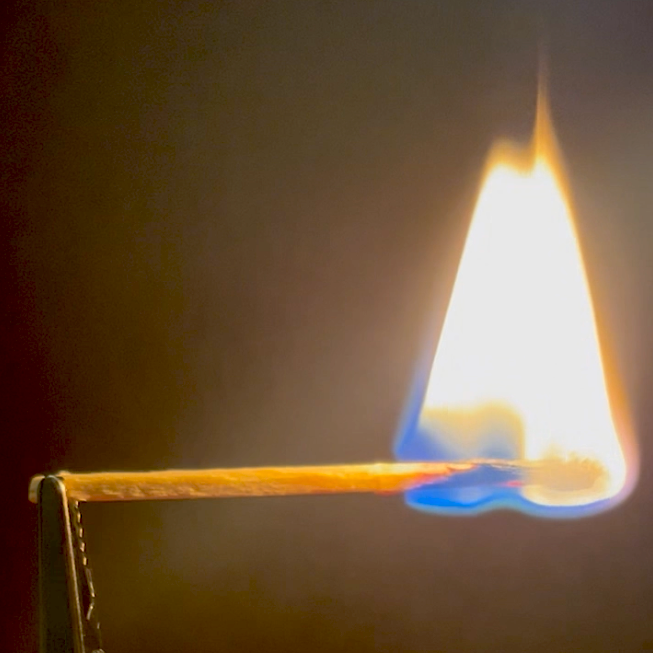}
         \caption{Left: Frame 131}
    \end{subfigure}
    \vspace{0.5em}
    \begin{subfigure}[h]{\columnwidth}
         \centering
         \includegraphics[width=0.63\columnwidth]{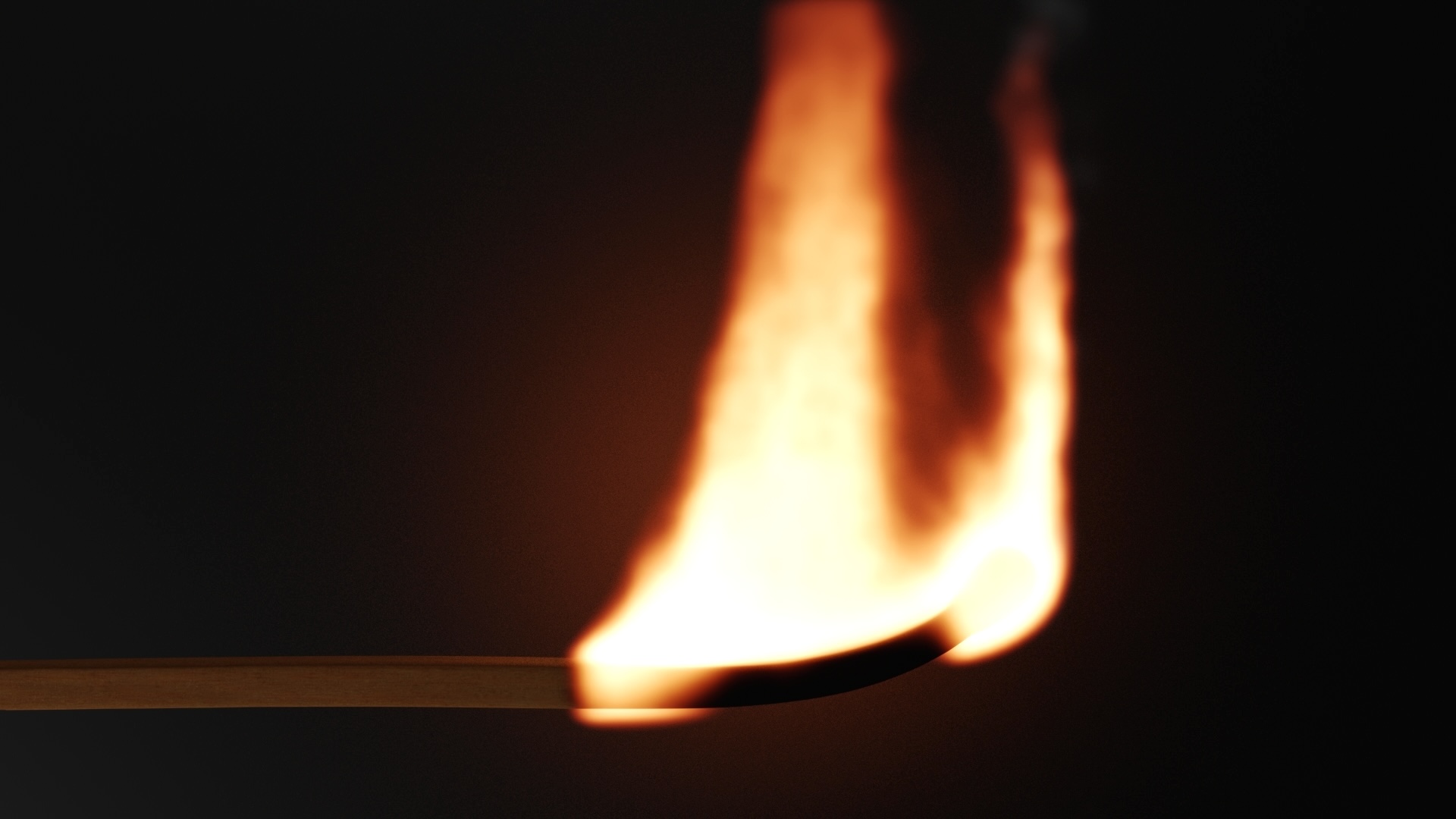} \includegraphics[width=0.355\columnwidth]{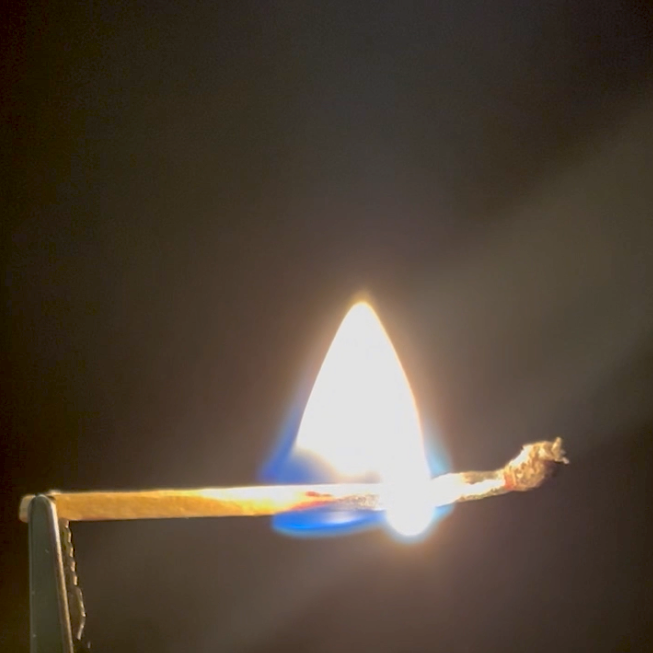}
         \caption{Left: Frame 184}
    \end{subfigure}
    \vspace{0.5em}
    \begin{subfigure}[h]{\columnwidth}
         \centering
         \includegraphics[width=0.63\columnwidth]{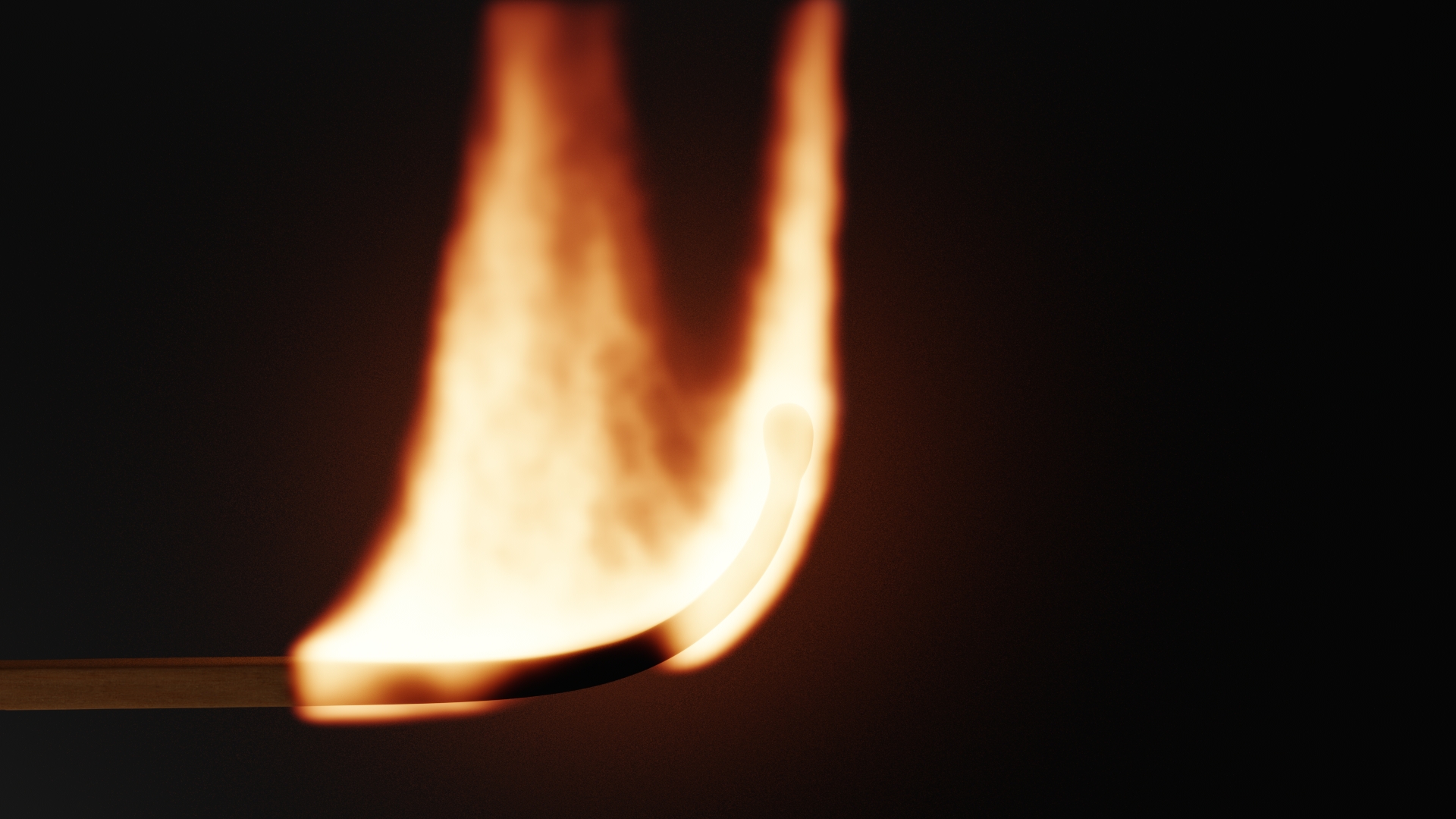} \includegraphics[width=0.355\columnwidth]{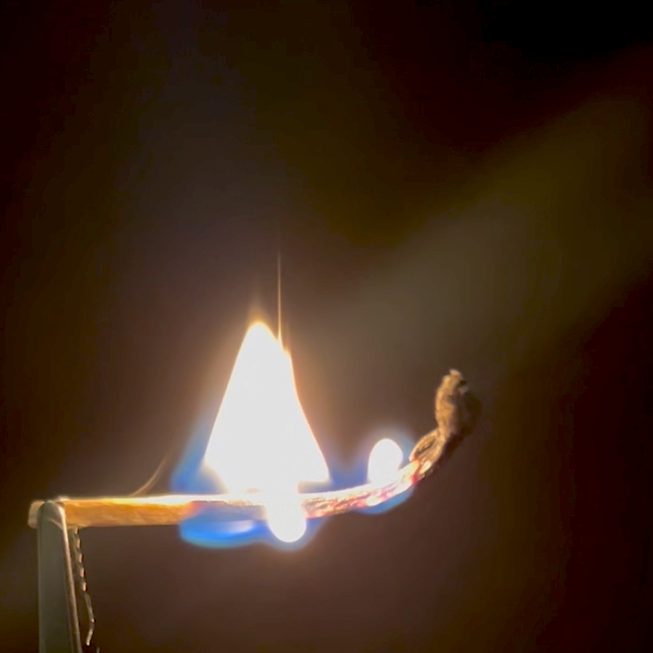}
         \caption{Left: Frame 237}
    \end{subfigure}
    \begin{subfigure}[h]{\columnwidth}
         \centering
         \includegraphics[width=0.63\columnwidth]{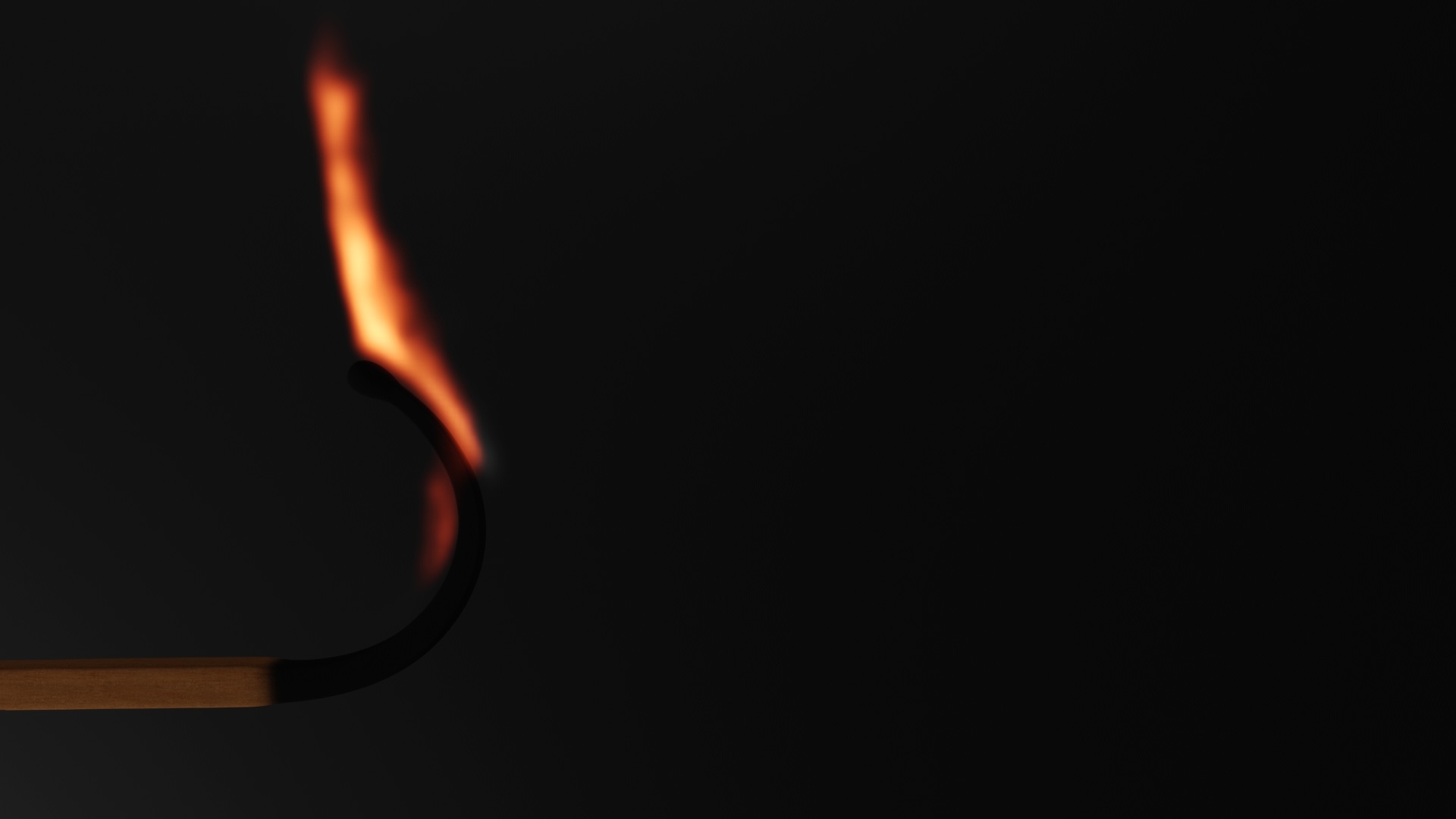} \includegraphics[width=0.355\columnwidth]{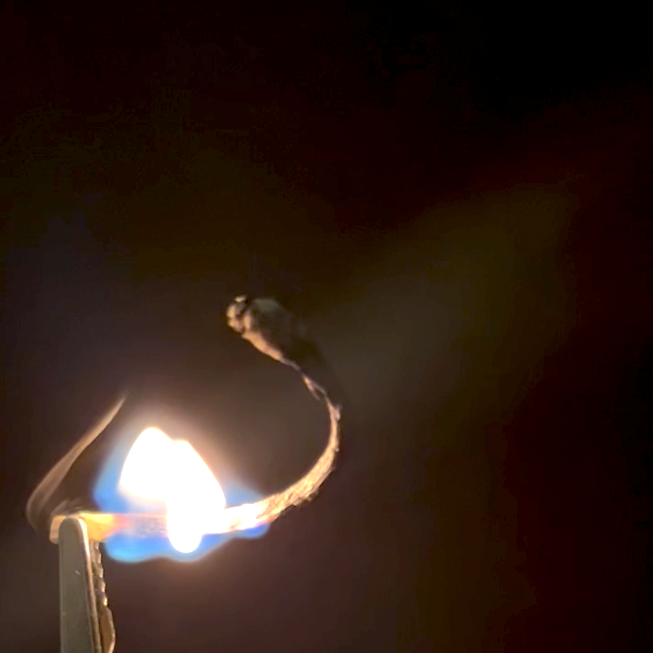}
         \caption{Left: Frame 338}
    \end{subfigure}
    \caption{(Left) Our simulated match curls as it burns. (Right) Comparable photos from our experiment burning a real match. Our simulation achieves qualitatively similar behavior in the deformation of the burning match.}
    \label{fig:burning_match}
\end{figure}

Lastly, Figure \ref{fig:burning_match} showcases a higher-resolution example, where we ignite the tip of a three-dimensional match. 
We also compare with our experiment of burning a real match.
The burning slowly spreads through the match.
Temperature rises due to buoyancy, and eventually the match becomes hotter on top of its surface versus on the bottom of its surface.
We apply isotropic shrinking by multiplying $\FF_p$ of particles whose temperature is above an evaporation temperature threshold by $1 + c \Delta t$, where $c$ is dependent on $c_{\text{shrink}} = 1.01$ as in Equation \eqref{shrink_scale}.
This shrinking causes the match to curl due to the temperature gradient on the match.
We use the burning status and fuel level of the particles to model the darkening of the texture.
The qualitative behavior of the match in our simulation agrees well with the solid deformation observed in the real-world footage.

\section{Conclusions and Future Work}

Our method hybridizes MPM and an incompressible fluid solver to produce dynamic and visually plausible thermodynamic simulation.
Our approach is directable with controls for flame front propagation speed and constitutive model changes over solid deformation and mesh-based fracture.
However there are a few obvious aspects that can be improved.
It would be natural to add thermomechanical phase changes into liquids, such as the melting wax examples of \citet{chen:2021:momentum}.
Our ignition method could also be extended to burning thin objects such as paper and rope; however, this would require different treatment of the fluid boundary conditions and constitutive model for the solid.
We only utilized one-way coupling, with the solid serving as the driving boundary conditions for the fluid. 
A treatment for two-way coupling such as in \citet{tampubolon:2017:wetsand} could enhance realism. 
Also, we utilize a voxelized incompressible fluid solver, which could benefit from implementing a cut-cell or sub-grid approach as in, e.g., \citet{bridson2015fluid,hyde:2019:unified}. Moreover, our method is constrained by the use of incompressible fluids for fire modeling, whereas incorporating a more advanced gaseous combustion solver as in \citet{nielsen2022physics} would yield more realistic results.
Further avenues for improvement include exploring more accurate heat exchange between solid and fluid materials, energy-conserving heat transfer from solid to fluid, and implementing multiple temperature fields per phase to enforce a sharp interface between the combustible solid and fluid.

\label{sec:conclusions}

% DO NOT INCLUDE ACKNOWLEDGMENTS IN AN ANONYMOUS SUBMISSION TO SIGGRAPH 2019
%\begin{acks}
%
%The authors would like to thank Dr. Maura Turolla of Telecom
%Italia for providing specifications about the application scenario.
%
%The work is supported by the \grantsponsor{GS501100001809}{National
%  Natural Science Foundation of
%  China}{http://dx.doi.org/10.13039/501100001809} under Grant
%No.:~\grantnum{GS501100001809}{61273304\_a}
%and~\grantnum[http://www.nnsf.cn/youngscientists]{GS501100001809}{Young
%  Scientists' Support Program}.
%
%
%\end{acks}

% Bibliography
\bibliographystyle{ACM-Reference-Format}
\bibliography{paper}

% Appendix
% \appendix
% \section{Switching Times}

% In this appendix,

\clearpage

%\begin{figure}[!ht]
 %   \centering
 %   \begin{subfigure}[h]{0.49\columnwidth}
 %        \centering
 %        \includegraphics[width=\columnwidth]{figs/debug_log_0.png} 
 %        \caption{Frame 0\vspace{0.5em}}
 %   \end{subfigure}
 %   \begin{subfigure}[h]{0.49\columnwidth}
 %        \centering
 %        \includegraphics[width=\columnwidth]{figs/debug_log_300.png}
 %        \caption{Frame 300\vspace{0.5em}}
 %   \end{subfigure}
 %   \begin{subfigure}[h]{0.49\columnwidth}
 %        \centering
 %        \includegraphics[width=\columnwidth]{figs/debug_log_500.png}
 %        \caption{Frame 500\vspace{0.5em}}
 %   \end{subfigure}
  %  \begin{subfigure}[h]{0.49\columnwidth}
  %       \centering
  %       \includegraphics[width=\columnwidth]{figs/debug_log_600.png}
%         \caption{Frame 600\vspace{0.5em}}
 %   \end{subfigure}
  %  \caption{Some views of the log example focusing on the shrinking and fracturing of the log.  Yellow indicates internal faces of the log mesh that become exposes due to mesh-based fracture as the log shrinks and deforms.}
  %  \label{fig:debug_log}
%\end{figure}

\clearpage

\newpage

\end{document}